# Towards a realistic noise modelling of quantum sensors for future satellite gravity missions

João Encarnação[a] J.G.deTeixeiradaEncarnacao@tudelft.nl,
Christian Siemes[a] C.Siemes@tudelft.nl,
Ilias Daras[b] Ilias.Daras@esa.int,
Olivier Carraz[b], olivier.carraz@esa.int,
Aaron Strangfeld[b] Aaron.Strangfeld@esa.int,
Philipp Zingerle[c] zingerle@tum.de,
Roland Pail[c] roland.pail@tum.de
[a] Delft University of Technology, Kluyverweg 1, 2629 HS, Delft, the Netherlands
[b] European Space Research and Technology Centre, Keplerlaan 1, 2200 AG Noordwijk, the Netherlands
[c] Technical University of Munich, Arcisstraße 21, D-80333 Munich, Germany

## Abstract

Mapping the Earth's gravity field from space offers valuable insights into climate change, hydro- and biosphere evolution, and seismic activity. Current satellite gravimetry missions have demonstrated the utility of gravity data in understanding global mass transport phenomena, climate dynamics, and geological processes. However, state-of-the-art measurement techniques face noise and long-term drift limitations, which propagate into the recovery of Earth's time-varying gravity field. Quantum sensors, particularly Cold Atom Interferometry (CAI), offer promise for improving the accuracy and stability of space-based gravity measurements. Therefore, CAI has emerged as a promising measurement technique for future gravimetric satellite missions due to their potential for measuring gravitational forces and gradients with high precision and accuracy, particularly at low frequencies (sub-mHz). This study explores the sensitivity of CAI accelerometers and gradiometers to the errors in measuring the satellite's attitude. We explore the low-low satellite-to-satellite and gravity gradiometry concepts and build the respective analytical models of measurements and associated errors. We selected an ambitious scenario for CAI parameters that illustrates a potential path for increasing instrument accuracies and capabilities for space gravimetry. Two operational modes, concurrent (where a new cloud is generated while another is moved to the interferometric chamber) and sequential (where cloud generation and interferometry happen in the same place), are compared to mitigate the effects of inaccurately known attitude rates on Coriolis accelerations. The sequential mode shows the potential to reduce these effects since the atom cloud has an initial zero velocity. Otherwise, the Coriolis effects are dominant in the concurrent operational mode. We additionally consider the impact on attitude uncertainty in the context of errors related to the reference frame rotation from the body to the Earth's co-rotating frames. The CAI configuration considered in this study allows for the time variable gravity signal to be observed in the case of low-low Satellite-to-Satellite Tracking missions. Still, it is insufficient for gravity gradient missions because of the reduced signal amplitude. We find it essential to understand and navigate the inherent technical challenges associated with quantum sensors to secure an efficient path towards exploiting this technology to monitor the gravity field changes.

# 1. Introduction

Recent advancements in quantum technology (Belenchia et al., 2022; Benyoucef, 2025) have spurred interest in developing highly sensitive instruments for gravitational measurements from space. Among these instruments, Cold Atom Interferometry (CAI) accelerometers and gradiometers are promising due to their potential for observing gravitational forces and gradients with high precision and accuracy. The microgravity environment in space allows for substantially longer interrogation time than on the ground, which is currently a limitation in developing these instruments.

The European Commission, backed by several European countries, has recently announced substantial investments in quantum technology to tackle contemporary digital challenges, including secure communication and computing power. Europe has been at the forefront of spaceborne quantum sensor development since the early 2000s, with initiatives such as the *Interférométrie atomique à sources Cohérentes pour l'Espace* (ICE) and *Matter-Wave Interferometry in Weightlessness* (MAIUS) advancing the field. The development of quantum sensors requires close collaboration between academia, industry, and space agencies to overcome technological challenges and prepare for future space missions (Kaltenbaek et al., 2021). Furthermore, integrating quantum sensors in space missions might enhance Earth observation and facilitate the exploration of celestial bodies like the Moon and Mars, ushering in a new era of space science and technology (Abend et al., 2023).

Mapping the Earth's gravity field from space offers unique insights into climate change, hydro and cryosphere evolution (Groh et al., 2019; Jiang et al., 2014), and seismic activity monitoring and modelling (Han, 2006), amongst others. Past satellite gravimetry missions, such as NASA and DLR's Gravity Recovery and Climate Experiment (GRACE), NASA and GFZ Helmholtz Centre for Geosciences' GRACE Follow-On (GRACE-FO) and ESA's Gravity Field and Steady-State Ocean Circulation Explorer (GOCE), have demonstrated the importance and relevance of gravity data in understanding global mass transport phenomena, climate dynamics, and geological processes. However, classical measurement techniques face limitations regarding noise and drift, which hinder the recovery of Earth's time-varying gravity field. ESA's Next Generation Gravity Mission (NGGM) and NASA and DLR's GRACE-Continuity (GRACE-C) mission are poised to leverage classical sensors to enhance Earth observation capabilities (Massotti et al., 2021; Cesare et al., 2022; Heller-Kaikov et al., 2023). The advent of quantum sensors, particularly CAI technology, holds promise for improving the accuracy and stability of space-based gravity measurements. Unlike some studies of spaceborne CAI (e.g., Zhu et al., 2022), we focus on the requirements for resolving the Atmosphere, Ocean, Hydrology, Ice and Solid-Earth (AOHIS) components of the time-variable gravity field model (Dobslaw et al., 2016).

While we are not proposing any specific mission configuration, we consider two measurement concepts: changes in the range between two satellites orbiting in formation, a technique called low-low Satellite-to-Satellite (ll-SST), and in situ gravity gradients collected by a gradiometer onboard a single satellite, called Gravity Gradiometry (GG). In the case of ll-SST, the changes in the ISR, which is the *main observable*, are caused by gravity and non-gravitational accelerations affecting the motion of the satellites. This measurement concept requires accelerometers that are, by design, sensitive to non-gravitational accelerations, allowing the gravitational accelerations to be isolated. In the case of GG, the gradiometer comprises one or more pairs of accelerometers, and their differential acceleration measurements, which are the *main observable here*, are related to the gravity gradient at the location of the satellite. We designate the accelerometers onboard current and past gravimetric missions by electrostatic instruments, cf. Section 2.3.1 and quantum instruments are further discussed in Section 2.3.2. We only consider these measurement concepts because we assume they could be realised within the next two decades. It is in this period that CAI technology will most likely mature into a primary sensor for collecting non-gravitational or gravity gradient observations. In this context, the informed decision of which measurement concept is best suited to CAI technology, given current and foreseeable technological limitations of all relevant on-board sensors, is an important aspect to clarify.

Associated with the measured physical quantities of these measurement concepts are their respective *product noise*, whose quantification in the spectral domain is the focus of this study. It is a prediction of the accuracy of the actual gravimetric observation that is exploited to produce a gravity field solution, but excluding the errors in models of 1) well-known gravity signals such as tidal accelerations and, 2) high-frequency signals that cannot be measured, such as non-tidal atmospheric and ocean gravity variations, also known as temporal aliasing. We do not consider any model errors in this study, because those have the same magnitude irrespective of the gravimetric sensor accuracy. The product noise is different to the aforementioned main observables, i.e., ISR for ll-SST and differential accelerations for GG. The main observables must be processed to isolate the gravity signal, to which other instruments are critical, notably the attitude and position relative to Earth. Therefore, the accuracy of the product noise is a function of the sensitivity of all relevant sensors and auxiliary instruments, not only of the instrument collecting the main observable.

To retain a clear focus on the instrument design and to limit the extent, we perform no further analyses of the impact of the developed instrument noise models on concrete satellite gravity missions. For that, we refer to Zingerle et al. (2024), who present the propagation of the predicted noise spectra to the gravity field coefficients. That study puts the sensor errors in context with other sources of errors that we are unable to consider, notably model errors. It presents a broad collection of possible future satellite gravity constellation scenarios. Both studies are part of the same ESA QSG4EMT project; see acknowledgements. The results presented in this study are part of the sensitivity analysis of quantum instrument performances, which was fed into the trade space analysis of Quantum Space Gravimetry (QSG) mission architectures in conjunction with a consolidated user requirements database for future QSG missions.

We describe the errors with Amplitude Spectral Density (ASD), which is the square root of the Power Spectral Density (PSD), to which we associate the symbol $s$ in the relevant equations. We start with the listing of instrument noise in Section 2, to collect existing or expected error spectra of all relevant sensors, i.e., attitude (Section 2.1), ISR (Section 2.2), classical and quantum accelerometers (Section 2.3) and discuss the consequences to GG and motivate why we do not consider classical GG (Section 2.4). With a suitable collection of error spectra, we proceed with describing how to quantify the product noise in Section 3, for ll-SST (Section 3.1) and GG (Section 3.2). We describe our model for the errors in the quantum accelerometers in Section 3.1.2.1 and how the Coriolis accelerations affect it in Section 3.1.2.2, in the context of ll-SST. In Sections 3.2.2.1 and 3.2.2.2, we expand this discussion to GG, regarding errors in the CAI gradiometer and rotational effects, respectively. In Section 4, we describe the impact of imperfect attitude reconstruction in the product noise of ll-SST (Section 4.1) and GG (Section 4.2). Section 5 applies what has been discussed so far to demonstrate the importance of contemplating the frame acceleration errors in Section 5.1. We quantify the full error spectra for classical (Section 5.2.1) and quantum (Section 5.2.2) accelerometers for ll-SST, and quantify the challenges with CAI GG in Section 5.3.

One limitation of previous studies is that they ignore the spectral properties of the errors (e.g., Stray et al., 2025). Although that may be the case for the quantum projection noise, the attitude sensors needed to measure and remove the frame rotation effects do not have white noise. It is of critical importance to include the relevant part of the spectrum, from 0.2 mHz related to orbital frequency to 20 mHz (preferably higher), related to spherical harmonic (SH) degree 113. Other studies considered the spectral amplitude of the estimated errors, but they were produced assuming white noise for the angular rotation errors (Romeshkani et al., 2025; HosseiniArani et al., 2025). The topic of CAI satellite gradiometry under numerous relevant scenarios and in good detail has been analysed by Mu et al. (2024) but without considering the effect of attitude errors in the Coriolis accelerations. Some studies focus on the (closed-loop) error propagation of assumed CAI accelerometers or the hybrid of electrostatic and CAI accelerometers to gravity field parameters, with special emphasis on the critical effect of temporal aliasing, without explicitly including frame effects (e.g., Abrykosov et al., 2019; Luthcke et al., 2021). The impact of frame accelerations on the phase measurements, which manifests as a loss of interferometer contrast, has been analysed in Beaufils et al. (2023), where it was shown not to be significant relative to the quantum projection noise. We note that this is not the same as what we present in this study, since we focus on the frame accelerations acting on the atom cloud and how they are inaccurately reconstructed from imperfect attitude sensors.

This study intends primarily to evaluate the effect of current attitude measurement technology on the gravimetric observations collected with CAI. More specifically, we investigate the impact of an inaccurately known attitude on the Coriolis accelerations and propose mitigation strategies. We also quantify the effect of reference frame transformations of the measurements in the presence of inaccurate attitude information. With this, we intend to lay the groundwork for future developments in quantum gravimetry and, consequently, future satellite missions. We do this with emphasis on the sensitivity of CAI instruments to the satellite's angular velocity and show how best to leverage advancements in laser metrology to capitalise on the predicted higher level of accuracy in quantum accelerometers.

## 2. Instrument specifications

The errors with which an instrument makes a measurement are intrinsic to its design, measurement principle and construction. These errors are often predicted as part of the instrument's specifications, but generally, they are validated with actual measurements. In case of instruments that do not yet exist, this error takes the form of a requirement, which reflects both the goal accuracy and how feasible it is. The measured quantities are usually not very useful by themselves, and additional instruments are needed to measure auxiliary observations to exploit the information content of the main instruments fully. One example of an auxiliary observation is attitude, which is required in order to project the accelerometer measurements to the Line-Of-Sight (LOS) direction, where the ISR is implicitly defined, in the case of ll-SST and to reduce the differential accelerometer measurements from frame accelerations in the case of GG.

We present in this section the noise ASD of all main instruments we consider in this study, including K-band ranging instrument (KBR), Laser Ranging Interferometer (LRI), electrostatic accelerometers, and predicted CAI instruments, as well as instruments that collected auxiliary observations, such as the star sensors, differential wavefront sensors (DWS) and laser gyroscopes. Without being specific to which ll-SST or GG measurement principle they are relevant to, we group these instruments in attitude, ranging, and electrostatic and quantum accelerometers.

We provide a non-exhaustive list of sensors that are representative of state-of-the-art technology. In the case of the ll-SST, we also offer the expected performance of the laser ranging instrument in the foreseeable future. We do the same for the classic accelerometer in the form of NGGM's "goal" requirement and the MicroSTAR instrument to fly on NGGM (cf. Section 2.3.1). Naturally, we consider quantum instruments that will provide sufficient improvement relative to classical instruments: Section 2.3.2 describes our error model, and Section 2.3.3 motivates the choice of the relevant CAI parameters. We do not consider future attitude instruments because we do not find substantial improvements in these instruments in the literature. We mitigate this aspect by quantifying the required accuracy to properly exploit the observations collected by the quantum instruments in Section 5.
Multiple sensors that measure the same quantity may be installed in future gravimetric missions. In that case, we assume that $m$ different instruments with error spectra are combined optimally, similar to the combination of attitude sensors by Stummer et al. (2011):

$$s^2(f) = \left( \sum_{m=1}^{M} s_m^{-2}(f) \right)^{-1} . \quad (1)$$

Here, $s_m(f)$ is the noise ASD of sensor m, and $s(f)$ is the noise ASD when all sensors are combined optimally. We use the symbol $s$ to identify the frequency-dependent errors, even if omitting the dependency in frequency.
Integration or differentiation of the noise ASD of a quantity requires the factor $2\pi f$ or its inverse, respectively. In the general case of the $n$-th derivative ($n > 0$) or n-th integral ($n < 0$), using Lagrange's notation for differentiation and antidifferentiation:

$$s\left(f^{(n)}\right) = (2\pi f)^n s(f). \quad (2)$$

## 2.1. Attitude Sensors

This section presents the noise ASD of the attitude instruments considered in the study. We present a wide selection of attitude instruments to assess what is technically feasible in the foreseeable future. Figure 1 shows an overview of the attitude noise ASDs, where all quantities have been converted to attitude (left) and angular rates (right) for convenience.

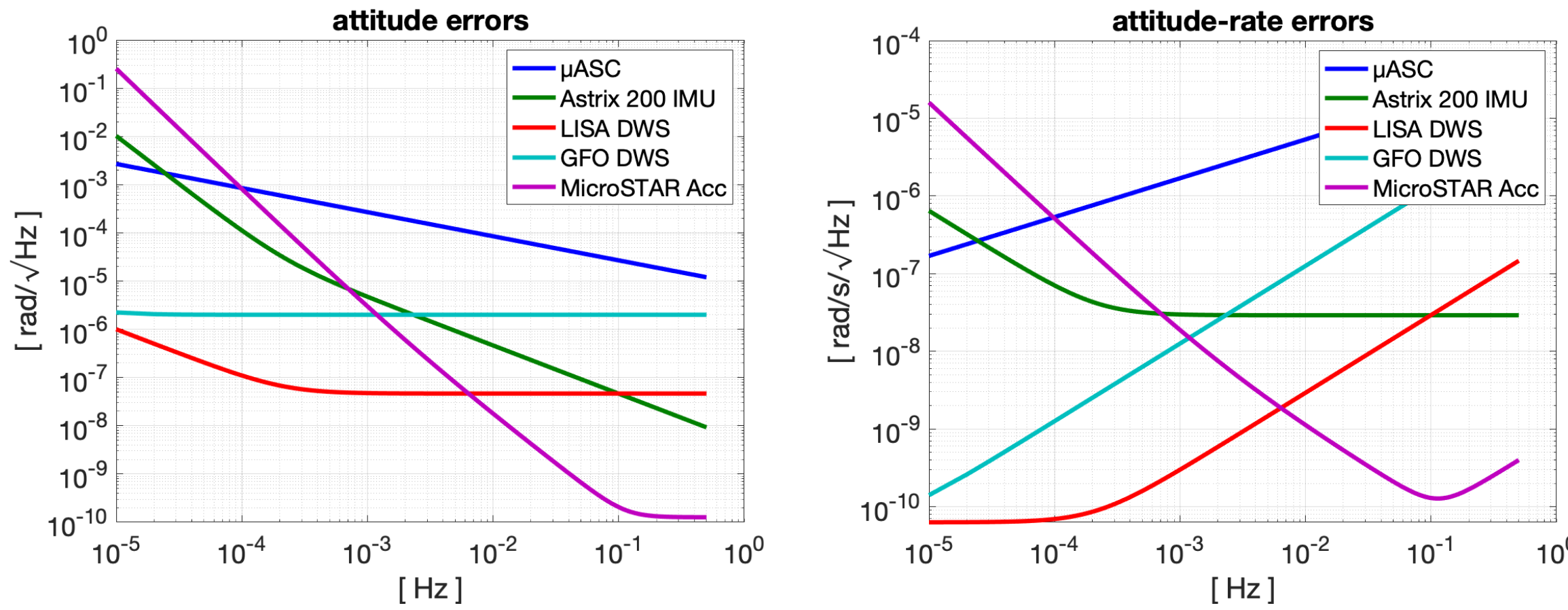

Figure 1: Overview of attitude errors in terms of attitude (left) and angular rates (right).

To clarify our notation, the symbols $s_\theta$, $s_\omega$ and $s_{\dot\omega}$ relate to angular, angular rate and angular acceleration errors, respectively.
The error spectra we consider are related to the accuracy with which the spacecraft's attitude is known in inertial space. Without going into detail, we assume that the attitude control and any necessary actuators are adequate to maintain the optimal operation of all instruments.

### 2.1.1. Star tracker

The star sensor of the Swarm satellites is the Micro Advanced Stellar Compass (μASC) (Herceg et al., 2017). We believe that this instrument is a representative state-of-the-art star sensor. (Goswami et al. 2021) analysed its in-flight accuracy and specified that the noise ASD as

$$s_{\mu\text{ASC},\theta}(f) = 8.5 \times 10^{-6}\sqrt{f^{-1}}\ \left[\frac{\text{rad}}{\sqrt{\text{Hz}}}\right]. \tag{3}$$

### 2.1.2. Inertial Measurement Unit

One of the most accurate inertial measurement units (IMU) is the Astrix 200 laser gyroscope, the accuracy of which is defined in its specification sheet (Airbus, 2022). It has a white noise component of $3 \times 10^{-8}$ rad/s (which happens to be twice as better as what HosseiniArani et al. (2025) considered) and a $f^{-1}$ component associated with a bias drift, as shown in Figure 2. The combined analytical expression is

$$s_{\text{IMU},\omega}(f) = 3 \times 10^{-8}\sqrt{1 + 4.6 \times 10^{-8} f^{-2}}\ \left[\frac{\text{rad/s}}{\sqrt{\text{Hz}}}\right]. \tag{4}$$

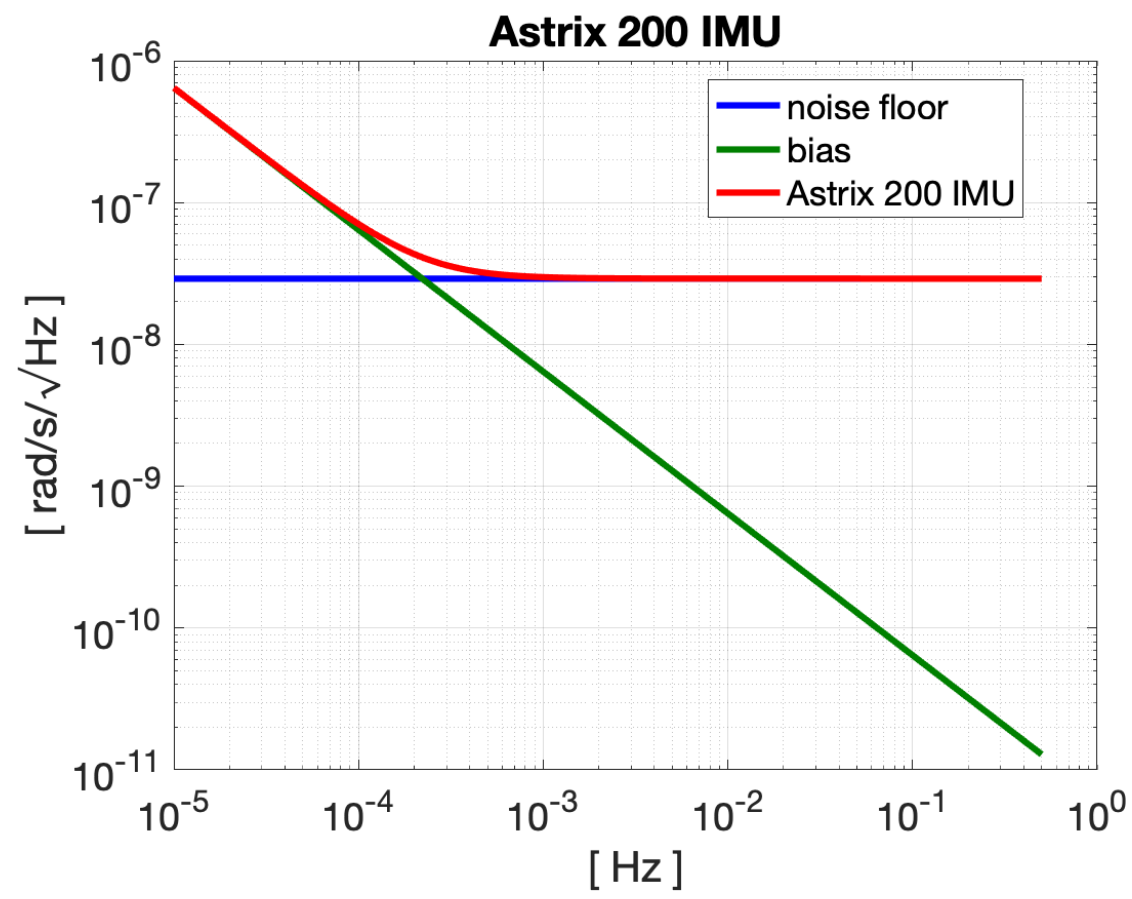

Figure 2: ASD of the angular velocity noise of the Astrix 200 IMU as inferred from the specification sheet (Airbus, 2022).

Although there are laser gyroscopes with accuracy at $3 \times 10^{-10}$ rad/s (Stedman et al., 2003), they are suitable for operation in space.

### 2.1.3. Differential Wave Sensor

The Differential Wave Sensor (DWS) measures pitch and yaw relative to the LRI. In combination with the GNSS-derived positions, it is possible to derive the absolute pitch and yaw attitude of the satellite because the GNSS positions provide the absolute attitude of the vector connecting the two satellites. The DWS noise spectra are provided for two cases in the following sections. As for the errors in GNSS $s_{\text{GNSS}}$ (supposing white noise with an amplitude of 1 cm), they proportionally affect the attitude error in the LOS unit vector $s_{\text{LOS},\theta}$. Finally, this effect is dampened proportionally to the inter-satellite distance $L_{\text{ISR}}$, assumed to be 200km:

$$s_{\text{LOS},\theta} = \frac{s_{\text{GNSS}}}{L_{\text{ISR}}} \cong 5 \times 10^{-8}\left[\frac{\text{rad}}{\sqrt{\text{Hz}}}\right]. \tag{5}$$

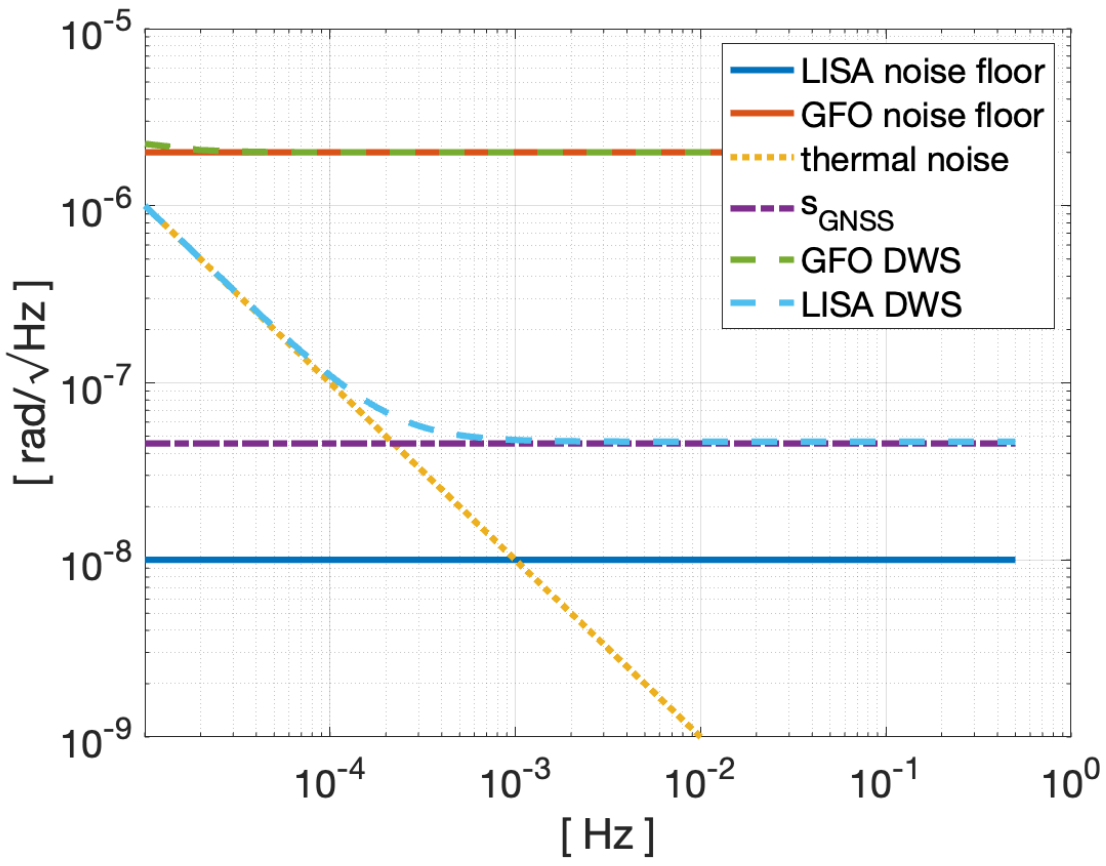

Figure 3: ASD of the attitude errors for the DWS of LISA (dashed blue line) (Schütze et al., 2013) and for the DWS of GRACE-FO (dashed green line) (Goswami et al., 2021).

#### 2.1.3.1. LISA

As shown in Figure 3, the noise ASD of the DWS is composed of the white noise floor (solid blue line) at the level of 10 nrad/$\sqrt{Hz}$ between 1mHz and 1 Hz and the thermal noise (dotted yellow line) with a spectrum of 1/f below 1 mHz (Schütze et al., 2013).

#### 2.1.3.2. GRACE-FO

Referring to Figure 3, for GRACE-FO, it is also possible to derive an estimate for the DWS-derived pitch and yaw attitude errors considering the white noise floor (red line), which has an amplitude of 2 μrad/$\sqrt{\text{Hz}}$ (Goswami et al., 2021). The thermal noise floor (dotted yellow line) and the attitude error of the LOS unit vector (dot-dashed purple line) are assumed to be equal to the LISA case in Section 2.1.3.1. The noise floor of the DWS sensor is dominant over the GNSS and the thermal components, in contrast to the DWS of LISA.

### 2.1.4. Accelerometer-derived attitude

Since each facet of the proof-mass cavity contains multiple electrodes, the MicroSTAR accelerometer can measure angular accelerations. The analytical expression for the error ASD is (Christoph Steiger et al., 2014):

$$s_{\text{MicroSTAR},\dot{\omega}}(f) = 1 \times 10^{-10}\sqrt{0.4 + 0.001f^{-1} + 2500f^4}\left[\frac{\text{rad}/s^2}{\sqrt{\text{Hz}}}\right]. \tag{6}$$

We assume this noise ASD is the same for all three axes, considering a cubic proof mass, identical gaps between the proof mass and electrodes on all sides, and neglecting the influence of the gold wire connected to the proof mass needed to neutralise the build-up of static charge.

## 2.2. Inter-Satellite Ranging

This section presents the noise ASD of the ISR instruments considered in the study. The overview of the noise ASD is shown in Figure 4.

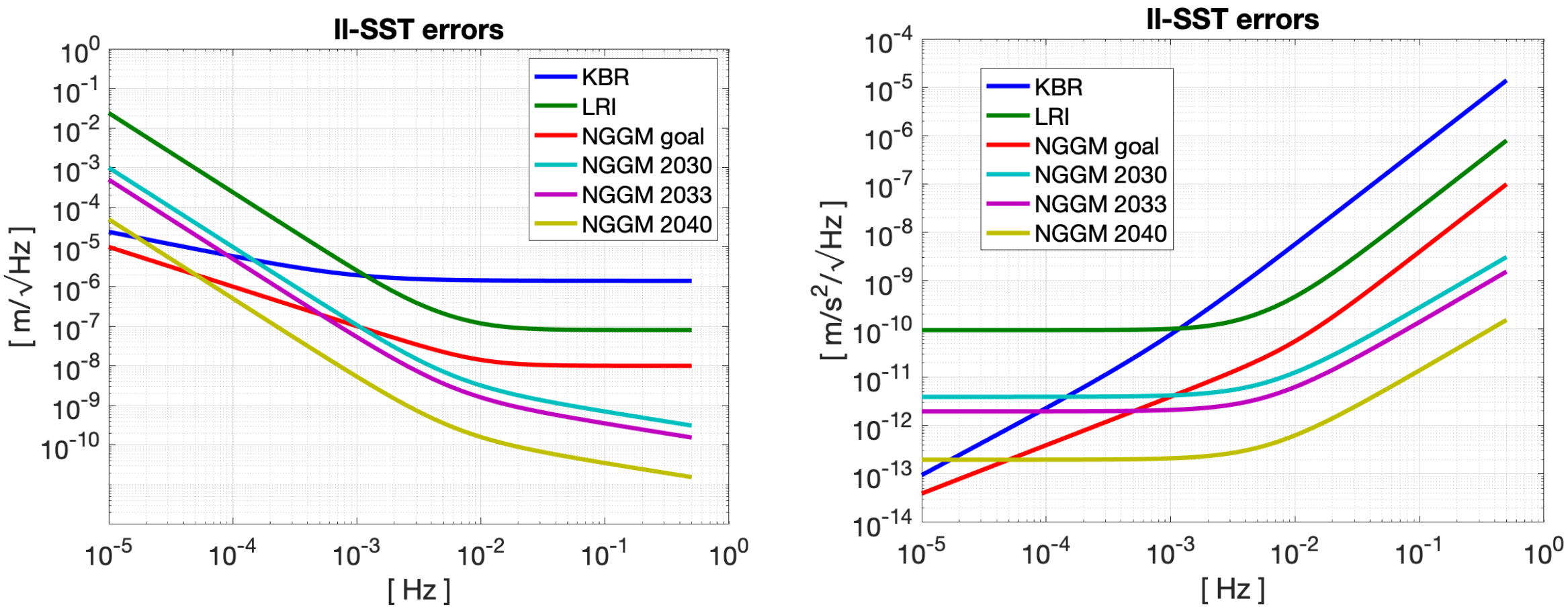

Figure 4: Overview of ISR noise ASD at the level of distance (left) and acceleration (right).

As we intend to quantify the errors for quantum gravimetric missions, which are currently in the early stages of development, we consider the error spectra associated with the *NGGM 2040* scenario (cf. Section 2.2.4). We report numerous other scenarios for the ranging instrument to contextualise our assumptions with existing instruments and assumptions in the literature.

### 2.2.1. GRACE-FO KBR

The KBR system is composed of thermal and Ultra-Stable Oscillator (USO)-related components (Sheard et al., 2012), illustrated in Figure 5 and described by the analytical expression:

$$\begin{aligned} s_{\mathrm{KBR},\rho} &= s_{\mathrm{KBR,thermal}} + s_{\mathrm{KBR,USO}}(f) \\ &= 1.4 \times 10^{-6} \left[\frac{m}{\sqrt{Hz}}\right] + 1.8 \times 10^{-8} f^{-\frac{1}{1.6}} \left[\frac{m}{\sqrt{Hz}}\right] \end{aligned} \qquad (7)$$

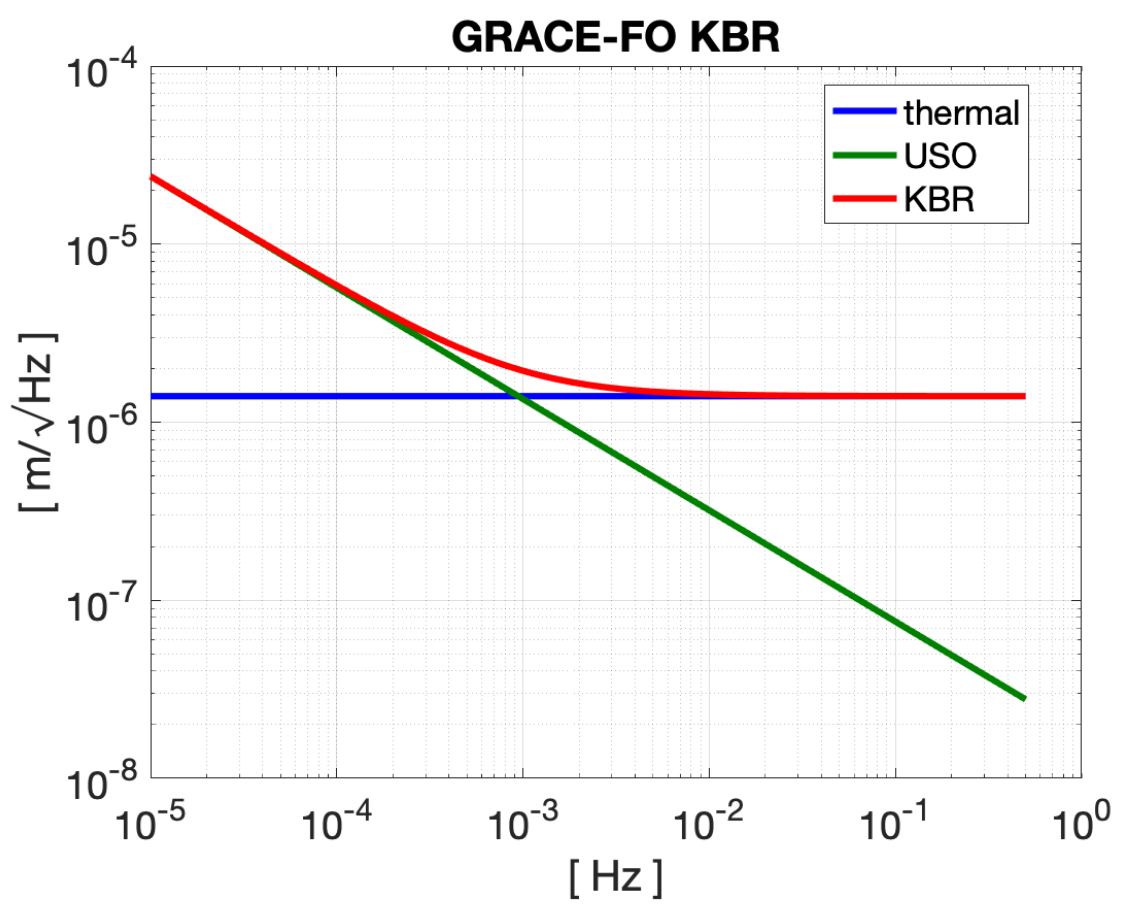

Figure 5: Noise ASD of the GRACE-FO KBR (Sheard et al., 2012, fig. 2).

### 2.2.2. GRACE-FO LRI

The noise ASD of the LRI of the GRACE-FO mission is given by the following analytical expression in terms of range noise (Kornfeld et al., 2019), shown as the green line in Figure 4:

$$s_{\mathrm{LRI},\rho}(f) = 8 \times 10^{-8} \sqrt{1 + (f/0.003)^{-2}} \sqrt{1 + (f/0.01)^{-2}} \quad \left[\frac{m}{\sqrt{Hz}}\right]. \qquad (8)$$

### 2.2.3. NGGM "goal" LTI

The ISR sensors presented so far are indicative of the existing instruments. A comparison between these and future QSG would not correctly represent the capabilities of the former. For that reason, we include in this study the "goal" performance of the NGGM mission concept (Massotti et al., 2021). The associated analytical expression is a function of the inter-satellite distance $L_{ISR}$ (cf. red line in Figure 4):

$$s_{\mathrm{NGGM},\rho}(f) = L_{\mathrm{ISR}} 10^{-13} \sqrt{1 + (0.01/f)^2} \sqrt{1 + (0.001/f)^2} \;\; \left[\frac{m}{\sqrt{Hz}}\right]. \qquad (9)$$

### 2.2.4. NGGM predicted

For the projected accuracy of future ISR laser instruments, we consider the following spectra, which are predicted to be representative of the errors of these instruments at different years (p.c. Vitali Müller, Albert-Einstein-Institut, Hannover, March 2023):

$$s_{\mathrm{NGGM\ 2030},\rho}(f) = L_{\mathrm{ISR}} \frac{\mathbf{1} \times 10^{-15}}{f} + \frac{\mathbf{1} \times 10^{-13}}{f^2} \;\; \left[\frac{m}{\sqrt{Hz}}\right], \qquad (10)$$

$$s_{\mathrm{NGGM\ 2033},\rho}(f) = L_{ISR} \frac{\mathbf{5} \times 10^{-16}}{f} + \frac{\mathbf{5} \times 10^{-14}}{f^2} \;\; \left[\frac{m}{\sqrt{Hz}}\right], \qquad (11)$$

$$s_{\mathrm{NGGM\ 2040},\rho}(f) = L_{ISR} \frac{\mathbf{5} \times 10^{-17}}{f} + \frac{\mathbf{5} \times 10^{-15}}{f^2} \;\; \left[\frac{m}{\sqrt{Hz}}\right]. \qquad (12)$$

The subscript indicates the year in which the instrument is predicted to be ready for flight. Refer to the teal, purple and yellow lines, respectively, in Figure 4. We note that Equation ( 10 ) is equivalent to Equation ( 9 ), considering only thermal noise at low frequency (<1mHz) and, therefore, no dependency on inter-satellite distance at low frequency. Equation ( 11 ) is an improvement of factor 2 over Equation( 10 ), as already shown in GRACE-FO (Abich et al., 2019). Equation ( 12 ) is one order of magnitude improvement, as expected by LISA (Dahl et al., 2019).

## 2.3. Accelerometry

### 2.3.1. Electrostatic accelerometry

As electrostatic accelerometers, we consider the goal requirements of the NGGM mission concept and the performance of the MicroSTAR accelerometer for the linear acceleration measurements. Figure 6 presents an overview.

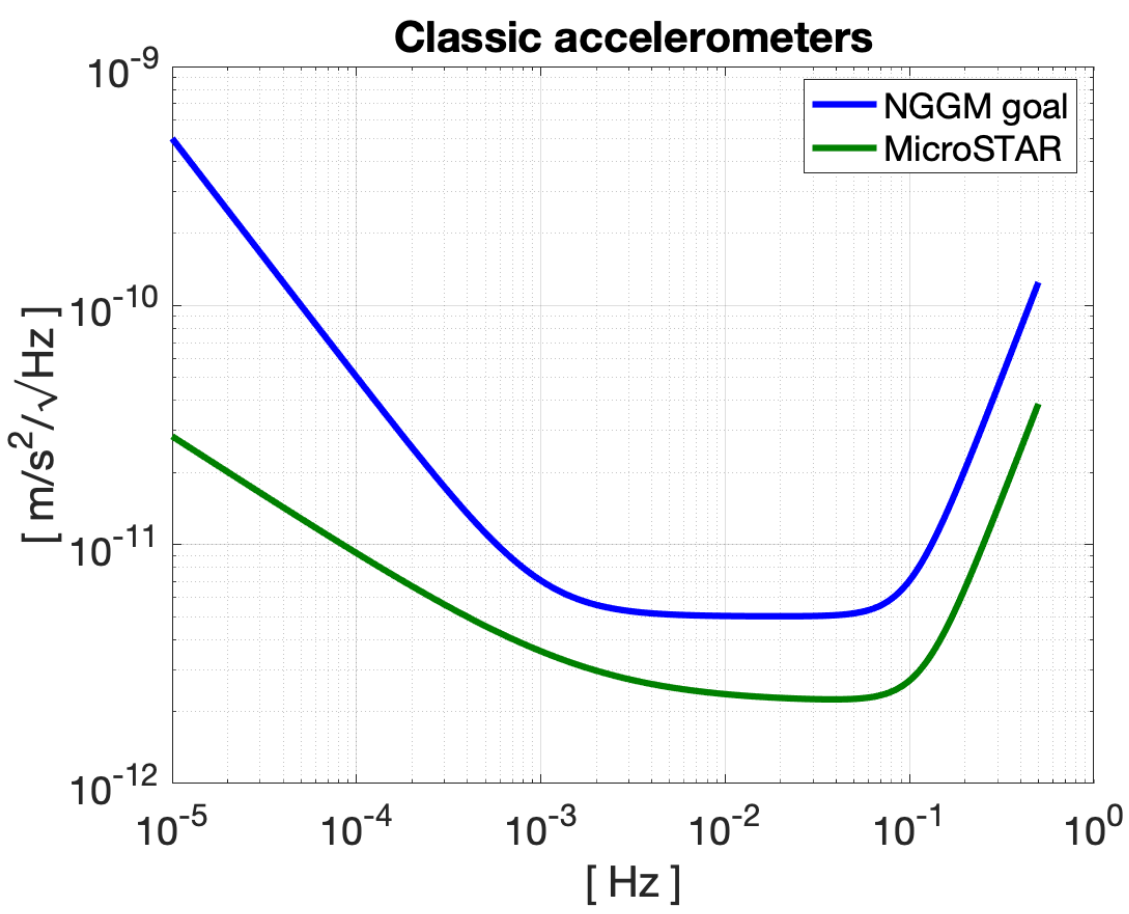

Figure 6: Overview of the noise ASD of electrostatic accelerometers.

#### 2.3.1.1. NGGM "goal" accelerometer

There are two requirements for the accelerometer performance of the NGGM mission concept: "goal" and "threshold" (Massotti et al., 2021). We selected the "goal" scenario with the associated noise ASD is defined by (cf. blue line in Figure 6):

$$s_{\mathrm{NGGM,ng}}(f) = 5 \times 10^{-12}\sqrt{1 + (0.001/f)^2 + (100f^2)^2}\ \left[\frac{\mathrm{m}/s^2}{\sqrt{\mathrm{Hz}}}\right] \tag{13}$$

#### 2.3.1.2. MicroSTAR

The MicroSTAR has a noise ASD given by the expression (Christophe et al., 2018), shown as the green line in Figure 6:

$$s_{\mathrm{MicroSTAR,ng}}(f) = 2 \times 10^{-12}\sqrt{1.2 + 0.002f^{-1} + 6000f^4}\ \left[\frac{\mathrm{m}/s^2}{\sqrt{\mathrm{Hz}}}\right]. \tag{14}$$

### 2.3.2. Quantum accelerometry

We assume the following CAI scheme (Malossi et al., 2010):

i) a Bose-Einstein Condensate (BEC) atomic cloud is produced from a Magneto-Optical Trap (MOT) by laser cooling and magnetic trapping techniques,
ii) a Raman pulse splits the wave-packet in two, through the transfer of momentum with the photons, kicking them in opposite directions along the axis of the Raman lasers and over the interrogation period $T$,
iii) a second Raman pulse imparts opposite momentum to the wave-packet, forcing them to converge,
iv) after the same period of interrogation, a third Raman pulse recombines the wave-packet, and
v) the interferometric measurement is conducted using the recombined wave-packet.

We use the term *wave-packet* to recognise the wave-particle duality of the BEC because, formally, there is no physical separation of the atomic clouds. Only the wave function is spread in two directions. Henceforth, the term *(atom) cloud* intends to loosely refer to both the physical cloud (in the case of quantum gradiometry in Sections 2.4 and 3.2.2) and the wave-packet, except when it is essential to make a distinction. The first and third pulses are also called π/2 pulses, and the second pulse is called a π pulse.

A non-zero acceleration $\boldsymbol{a}$ along the axis of the Raman laser induces a phase shift $\phi$ proportional to the acceleration the atom clouds have experienced during $2T$ (Peters et al., 1999):

$$\phi = \boldsymbol{k}_{\mathrm{eff}} \cdot \boldsymbol{a}T^2. \tag{15}$$

The magnitude of the effective wavevector $k_{\mathrm{eff}}$ is inversely proportional to the wavelength of the Raman laser $\lambda$. In the case of the double-diffraction scheme considered in this study, because of the direct and reflected Raman laser:

$$k_{\mathrm{eff}} = \frac{8\pi}{\lambda}. \tag{16}$$

By introducing the degree of entanglement $\alpha$ (e.g. for $\alpha = 0$, there is no entanglement and therefore reaching quantum projection noise; for $\alpha$=1 the Heisenberg limit is attained), with interferometer contrast C and the number of atoms $N$, we expand the simple relation dependent only on the SNR, which is proportional to $\sqrt{N}$ (Sorrentino et al., 2010) to model the interferometric phase noise as:

$$s_{\Phi} = \frac{1}{CN^{\frac{1+\alpha}{2}}}. \quad (17)$$

The interferometer contrast represents the intensity at which the interferometry fringes appear in the detector. The beam splitting efficiency and any external perturbations influence its value because they lead to a loss of atoms in the phase shift, mainly due to non-inertial effects. However, there are techniques to mitigate the rotation-induced loss of contrast (Kaczmarczuk et al., 2025). Experimental or assumed values range from 0.5 (Zahzam et al., 2022), 0.6 (Zhu et al., 2022) or 0.65 (Peters et al., 1999) to $\mathrm{C} = 0.8$ (Knabe et al., 2022), while the maximum value is 1 (Douch et al., 2018). The degree of entanglement refers to different quantum enhancement techniques that allow the phase difference after the interrogation time $T$ to be observed more accurately (Szigeti et al., 2021), with *Spin Squeezing* being the most common (Gross, 2012). Parameter $\alpha$ reflects the proportion of atoms in the cloud that are entangled, ranging from 0 to 1, where the value 0 means there is no entanglement. In the latter case, your error model is identical to Zahzam et al. (2022) and Zhu et al. (2022).

In the case of a concept that involves multiple momentum diffractions, the momentum transfer $\delta p$ is the product of the reduced Planck constant $\hbar$ with the magnitude of the wavevector $k_{\text{eff}}$ and the Momentum Space Separation $\beta$, which has a unit value for the baseline double diffraction:

$$\delta p = \hbar \beta k_{\text{eff}}, \quad (18)$$

and for a single-axis instrument where the effective wavevector is aligned in the measured acceleration $a$, Equation ( 15 ) can be written to accommodate different values for $\beta$ as

$$\phi = \frac{\delta p}{\hbar} a T^2. \quad (19)$$

Under these assumptions, the CAI accelerometer shot-to-shot (s2s) sensitivity, also called *quantum projection noise*, is:

$$s_{\text{CAI,ng}}^{(s2s)} = \frac{\hbar}{\delta p T^2} s_{\Phi} = \frac{1}{\beta k_{\text{eff}} T^2 C N^{\frac{1+\alpha}{2}}} \quad (20)$$

We assume that the noise spectra of the CAI accelerometers are flat, corresponding to white noise (Sorrentino et al., 2010); for this reason, the standard deviation is sufficient to describe these errors entirely. This characteristic is the main advantage over classical accelerometers (Migliaccio et al., 2019), which is dominated by noise at low frequencies resulting from bias thermal fluctuations and gold wire damping (Reubelt et al., 2014; Christophe et al., 2015, 2018), with a characteristic $f^{-n}, n \geq 1$ frequency dependence.

#### 2.3.2.1. Mode of operation

So far, we have restricted our analysis to the shot-to-shot sensitivity CAI sensitivity, which represents the best-case scenario where the measurements are made continuously without any interruptions. In reality, this is impossible because the atom cloud needs time to be prepared, which we assume to be $T_{\text{prep}} = 1s$ (Müntinga et al., 2013). Additionally, we define $T_{\text{cycle}}$ as the complete measurement cycle period.

We identify two distinct modes for the design and operation of the CAI:

- *Concurrent* atom cloud preparation and interrogation, where the interferometry takes place at the same time as the BEC is being prepared: $T_{\text{cycle}} = T_{\text{prep}}$
- *Sequential* atom cloud preparation and interrogation, the process for cloud preparation and interrogation do not overlap, leading to a more extended measurement cycle period: $T_{\text{cycle}} = 2T + T_{\text{prep}}$.

In the concurrent case, the next atom cloud can be launched before the CAI sequence of the current atom cloud is finished, i.e., the measurement cycle $T_{cycle}$ is only limited by the atom cloud preparation time $T_{\text{prep}}$, and we avoid any dead time between measurement cycles.
For both cases, the standard deviation of the CAI acceleration is:

$$s_{\text{CAI,ng}} = \sqrt{T_{\text{cycle}} s_{\text{CAI,ng}}^{(s2s)}}. \tag{21}$$

Additional considerations regarding how the Coriolis accelerations influence both operation modes are discussed in Section 5.1.
One consequence of the two operational modes is the cloud velocity, which is analysed in Section 5.1. The sequential mode of operation allows for an initial zero-atom cloud velocity since the preparation of the BEC and interferometric chambers is placed at the same location. The BEC is directly prepared at the CoM, and the wave packet propagates exclusively along the axis of the Raman laser (left of Figure 7), with momentum depicted by the blue arrows, resulting from the momentum transfer imparted by the Raman laser alone Zahzam et al., 2022. For the concurrent mode of operation, the atom cloud enters one side of the chamber with a non-zero (physical) velocity perpendicular to the Raman laser (green arrow in Figure 7), for example, $v_{\text{cloud}} = 2.5\ \text{cm/s}$ (Carraz et al., 2014; Trimeche et al., 2019). An additional recoil laser with an axis perpendicular to the Raman laser imparts the transverse velocity $v_{\text{cloud}}$ to the atom cloud. The first π/2 pulse is done on one side of the chamber by the first Raman laser, the π pulse is done by a second Raman laser in the middle of the chamber, and the third Raman laser on the opposite side of the chamber is responsible for the second π/2 pulse.

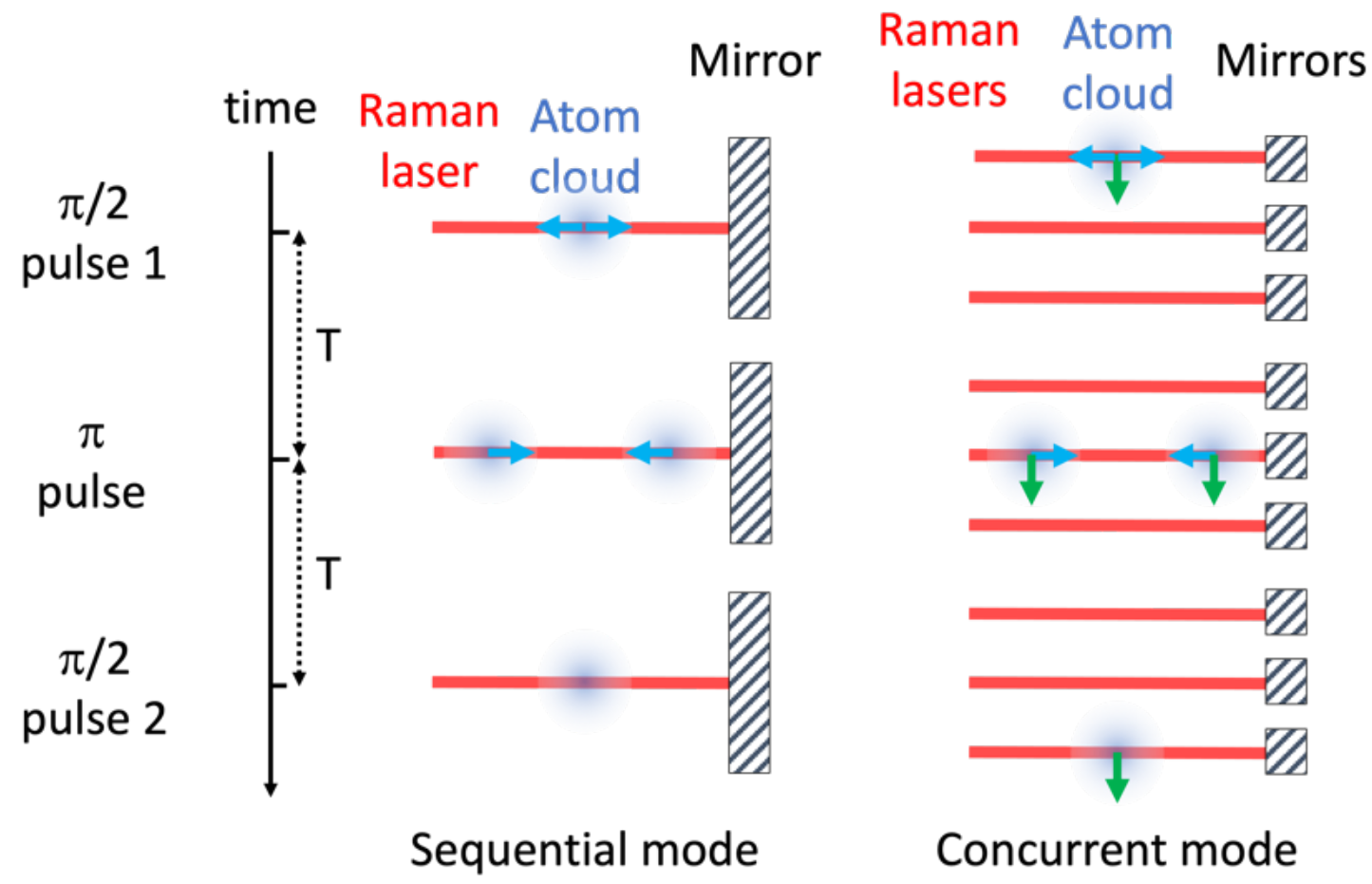

Figure 7: Diagram of the interferometry scheme for the sequential (left) and concurrent modes (right), showing the wave-packets (blue *Atom cloud*) after the respective Raman laser pulses, which add momentum as represented by the blue arrows. In the case of the concurrent mode (right), the transverse (physical) velocity represented by the green arrow is provided at the start of the measurement sequence by an additional laser (not shown) perpendicular to the Raman lasers.

For both operation models, the mirror(s) tilt to compensate for the satellite rotation, i.e., they rotate between the laser pulses of the CAI. The concurrent mode requires three mirrors in different locations along the direction of the atom's movement.

### 2.3.3. Selection of CAI parameters

To evaluate the utility of CAIs for low-low SST missions, we selected a set of representative instrument configurations based on a simplified performance model and heuristic cost proxies. These were used to identify parameter combinations that offer plausible sensitivity while remaining technically viable. Figure 8 illustrates the resulting parameter space, showing how performance varies against a qualitative cost metric. The selected configurations are not intended as optimal designs, but as reasonable inputs for further assessment within the broader study. Additional details of this analysis are shown in Appendix 8.7.

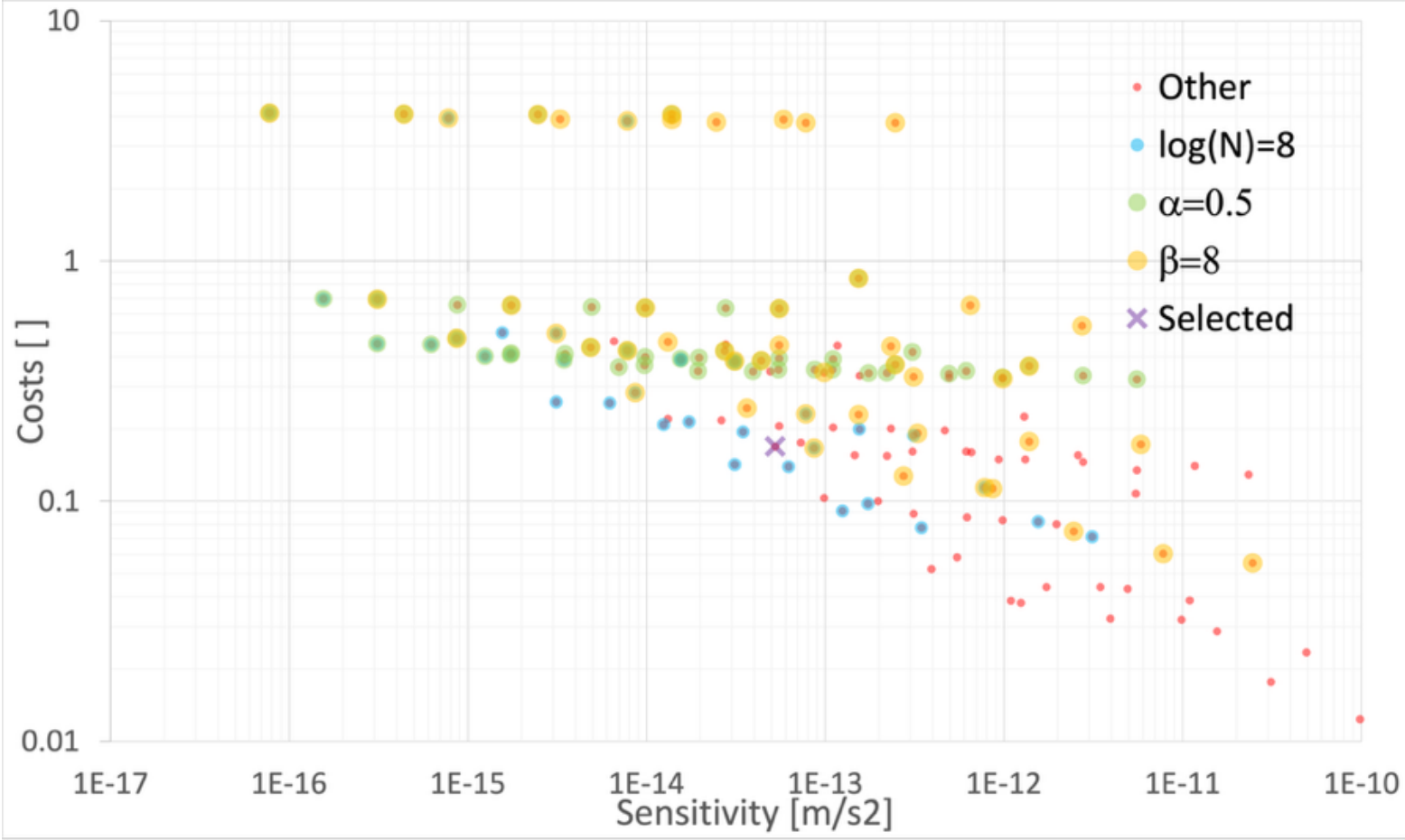

Figure 8: Trade Space Exploration, highlighting those options with extreme values for the CAI parameters (cyan, green and yellow) and the selection shown in Table 2 (purple cross).

## 2.4. Gradiometry

In both classical and quantum GG, we consider this technique to be implemented by combining pairs of accelerometers. As such, the error spectra of the electrostatic and quantum accelerometers can be directly converted to gradiometer measurements by dividing the former by the length of the gradiometer arm $L_{GG}$, which is the distance between pairs of accelerometers on the same axis. Therefore, considering the MicroSTAR accelerometer (Section 2.3.1.2), with a noise floor of $2 \times 10^{-12}$ $m/s^2$, a gradiometer built with these instruments with $L_{GG} = 0.5$m would have a noise floor of $4 \times 10^{-12}$ $s^{-2}$ or 4 mE.
The motivation for including a gradiometer in this study is to establish the requirements for a CAI gradiometer, considering that electrostatic accelerometers are unable to attain the necessary accuracy, as depicted in Figure 9. In this figure, the gravity gradient signal in the radial direction due to the time-variable gravity field, i.e. excluding the mean gravity field, at altitudes of 200 km and 300 km is shown in the yellow and red lines, respectively; the noise of GOCE gravity gradients is shown for reference in the blue line. Sub-mE gradiometry is required, which would need an electrostatic instrument that is at least one order of magnitude more accurate than the MicroSTAR accelerometer operating in ideal conditions.

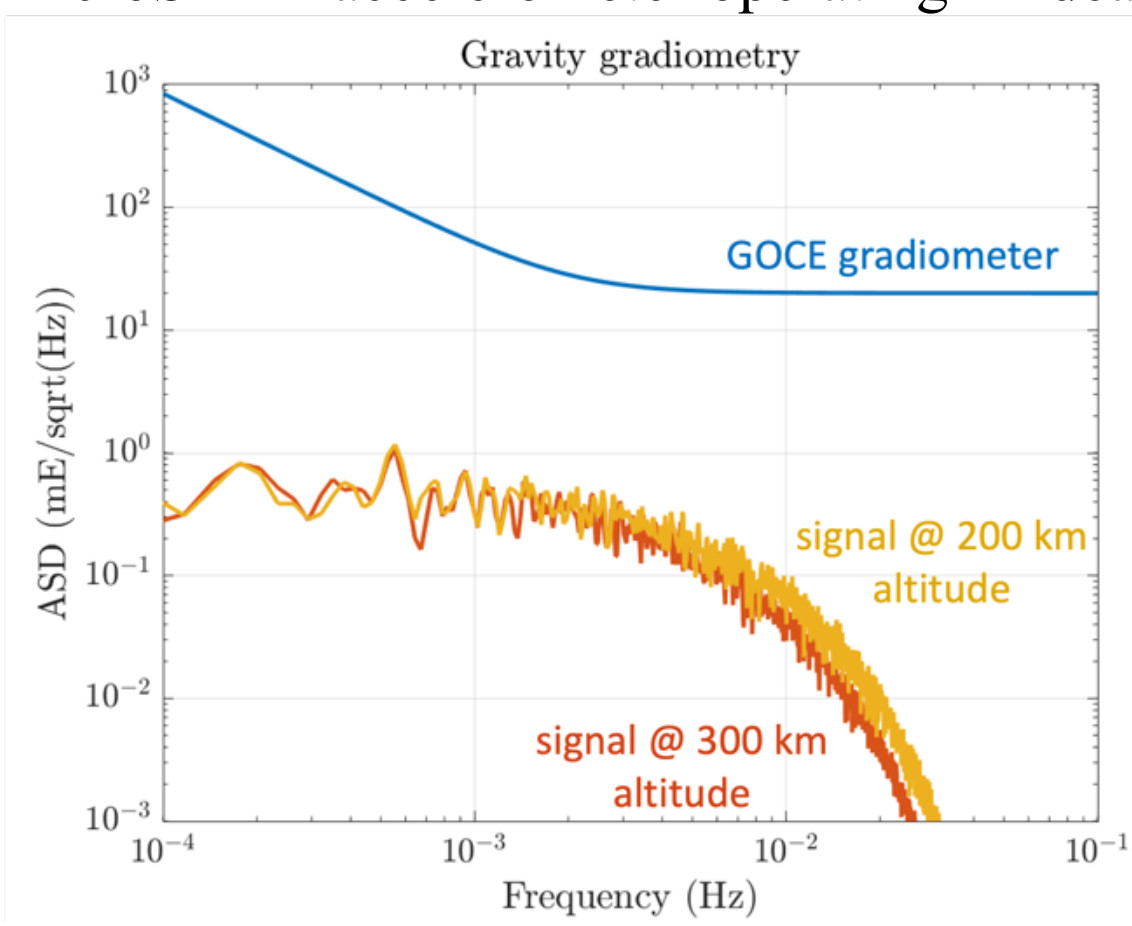

Figure 9: Comparison between the GOCE gradiometer errors (blue line) with the time-variable signal in terms of gravity gradient at 200 km (yellow line) and 300 km altitudes (red lines).

For quantum gradiometry, the measurement concept is similar to quantum accelerometry (Section 2.3.2), except that it is doubled along the axis where the gravity gradient is measured. The original atom cloud is physically split in two with a high-recoil laser pulse, originating two CAI accelerometers. Differential acceleration is therefore measured between the 2 CAI accelerometers. Common acceleration is then rejected as the 2 CAI share the same Raman laser pulses, and the gravity gradient along the Raman axis remains. This scheme applies to both sequential and concurrent modes of operation, with the obvious difference that the atom clouds are not moving perpendicularly to the Raman laser in the first mode. The amplitude of the gravity gradient as a function of the differential phase measurement and associated errors is closely related to $\phi$ in Equation ( 15 ) and $s_{\mathrm{CAI,ng}}$ in Equation ( 21 ), as discussed in Section 3.2.2.

# 3. Measurement concepts and noise modelling

We consider two measurement concepts: low-low satellite-to-satellite tracking (ll-SST) and GG. We discuss the most beneficial configurations when using electrostatic and quantum instrumentation for both measurement concepts. For all cases, we derive the product noise spectra as a function of the instrument noise spectra, which were defined in Section 2.
In the following, vector equations are transformed into their components using indices $i, j$ and $k$ for the coordinate axes x, y and z. Their relation is as arbitrary as the definition of the reference frames. For example, we may define $i, j, k \equiv x, y, z$ for along-track ll-SST and $i, j, k \equiv y, z, x$ for cross-track ll-SST, assuming the traditional axis nomenclature of $x$ being aligned with the along-track direction, $y$ with the cross-track direction and $z$ with the radial direction, in a circular low-Earth orbit (LEO) orbit.

## 3.1. ll-SST

The ll-SST concept relies on precise ranging between two satellites flying in the same low-altitude orbit, separated by a certain distance along the orbit (220 km in the case of the GRACE and GRACE-FO missions). In a variant of this concept, labelled cross-track ll-SST, the along-track distance is kept minimal, and the second orbit has a different right ascension of the ascending node than the first orbit to achieve ranging predominantly in the cross-track acceleration, away from the poles. The cross-track ll-SST has to respect a minimum along-track separation for collision avoidance at the poles, where the orbit crosses each other. In all ll-SST cases, the changes in the ISR are caused by variations in gravity and non-gravitational forces. Therefore, the concept foresees accelerometers to measure the non-gravitational accelerations so that the signal due to gravity can be extracted from the ranging measurements or drive a drag compensation system, similar to GOCE.

### 3.1.1. ll-SST with electrostatic accelerometers

The proposed ll-SST concept is illustrated in Figure 10. The ISR system is similar to GRACE-FO's LRI, which is implemented in the so-called racetrack configuration with a Triple Mirror Assembly (TMA). This concept has the benefit that the ISR is performed between the satellites' centre of mass (CoM) without physically occupying those locations while not compromising the ranging performance (Sheard et al., 2012). In the case of electrostatic instrumentation, we foresee two accelerometers, labelled ACC 1 and ACC 2, symmetrically placed around the satellite CoM. This arrangement allows the accelerometers to sense accelerations due to centrifugal and Euler forces and gravity gradients, which facilitates an accurate calibration of the accelerometers that would not be possible in the case of a single accelerometer placed into the satellite's CoM. Although GRACE and GRACE-FO, at least for a part of their missions, operated successfully with only one accelerometer at the CoM, the calibration process has always been problematic. It requires parameterisation strategies that are usually derived empirically (Teixeira da Encarnação et al., 2020). We assume that the two accelerometers are at the nominal distance of $L_{\mathrm{acc}}$ = 0.5 m from each other.

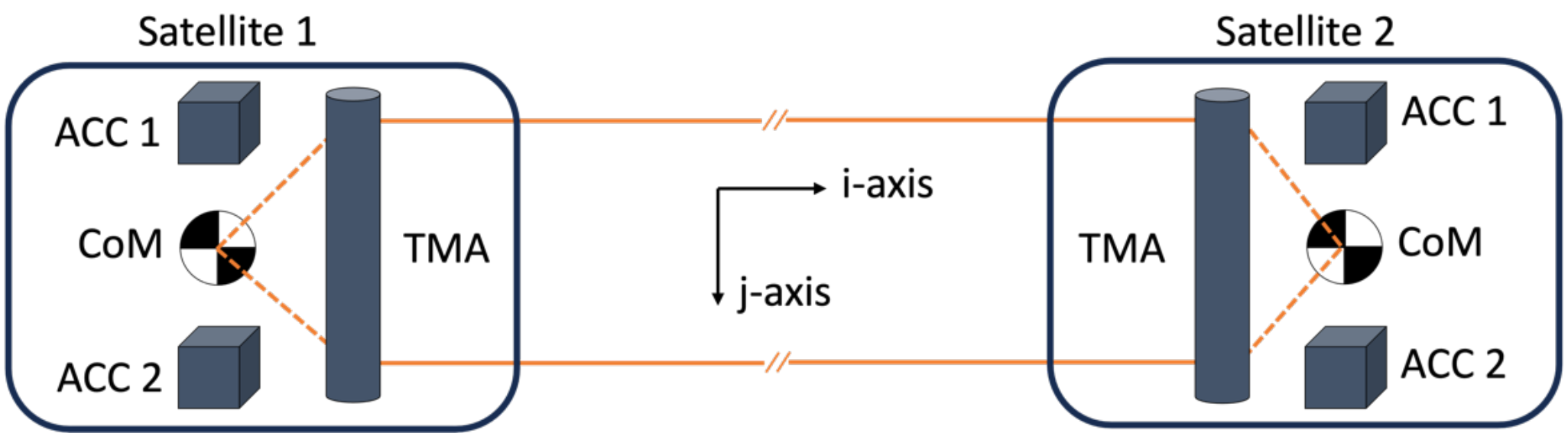

Figure 10: Diagram of the assumed concept for ll-SST with electrostatic accelerometers.

The LTI is equipped with a DWS sensor that measures the direction of the incoming laser beam. Combined with accurate knowledge of the positions of the satellites from GNSS, this allows for deriving the attitude of the satellites relative to the LOS vector (Section 2.1.3.2). Obviously, this provides pitch and yaw, but not the roll about the LOS. The attitude is also observed by star trackers (Section 2.1.1), augmented by the accelerometers (Section 2.1.4), and optionally, a high-performance IMU (Section 2.1.2).
To extract the non-gravitational acceleration from the measurements of the accelerometers, we form the so-called common-mode acceleration (Massotti et al., 2021). The common mode acceleration along the generic axis $i$ is:

$$a_{\mathrm{ng},i} = \frac{a_{1i} + a_{2i}}{2}, \tag{22}$$

where $i \equiv x$ is for the along-track ISR, and $i \equiv y$ is for the cross-track ISR. Through error propagation, we obtain the associated noise spectrum as a function of the linear acceleration error measured by a single accelerometer $s_{\mathrm{acc,ng}}(f)$ (Section 2.3.1):

$$s_{\mathrm{ng}}^2(f) = \frac{1}{2} s_{\mathrm{acc,ng}}^2(f). \tag{23}$$

The inter-satellite range acceleration $\ddot{\rho}$ contains the effects of both gravitational and non-gravitational accelerations acting on the two satellites, as well as the centrifugal acceleration of the LOS baseline (since this vector is rotating in inertial space). The non-gravitational accelerations are removed by subtracting the common-mode accelerations, assuming that the i-axis is aligned with the LTI axis. The centrifugal acceleration is a function of the ISR ρ and the inertial radial velocity $v_z$, assuming a nominal local horizontal attitude of the inter-satellite vector:

$$\ddot{\rho}_{\mathrm{grav}} = \ddot{\rho} - \frac{1}{\rho} v_z^2 - a_{\mathrm{ng},i}^{(1)} + a_{\mathrm{ng},i}^{(2)}. \tag{24}$$

Error propagation gives the noise spectrum of $\ddot{\rho}_{\mathrm{grav}}$ as a function of $\sigma_{\mathrm{ng}}$, the ISR error $\sigma_{\mathrm{ISR}}$ (Section 2.2) and errors in reconstructing the centrifugal accelerations $s_{\mathrm{c}}$:

$$s_{\ddot{\rho}}^2(f) = s_{\mathrm{ISR}}^2(f) + s_{\mathrm{c}}^2(f) + 2 s_{\mathrm{ng}}^2(f). \tag{25}$$

#### 3.1.1.1. Errors in reconstructing the centrifugal accelerations

The errors in reconstructing the centrifugal accelerations are given by (Ditmar et al., 2012):

$$s_c = -\frac{2v_z}{\rho} s_{v_z} + \frac{v_z^2}{\rho^2} s_\rho \approx -2\sqrt{\frac{GM}{R_{\mathrm{orb}}^3}} s_{v_z} + \frac{GM}{R_{\mathrm{orb}}^3} s_\rho. \tag{26}$$

Earth's gravitational parameter $GM$ is $3.986 \times 10^{14}\ \mathrm{m}^3/\mathrm{s}^2$ and $R_{\mathrm{orb}}$ is $6.897 \times 10^6$ m, which is related to an orbital period of 95 minutes, or 525.7 km altitude. The assumption for $R_{\mathrm{orb}}$ is relatively arbitrary for LEO, since the factor $GM/R_{\mathrm{orb}}^3$ only increases by 15.3% when the altitude decreases to 182 km. The assumptions for the magnitude of the radial velocity $v_z$ and associated errors $s_{v_z}$ are given in Appendix 8.6. Considering the error magnitude $s_\rho$ for both KBR and LRI ranging instruments, and the SuperSTAR accelerometers (cf. Section 2.2), Figure 11 contextualises the amplitude of $s_c$ for GRACE-FO.

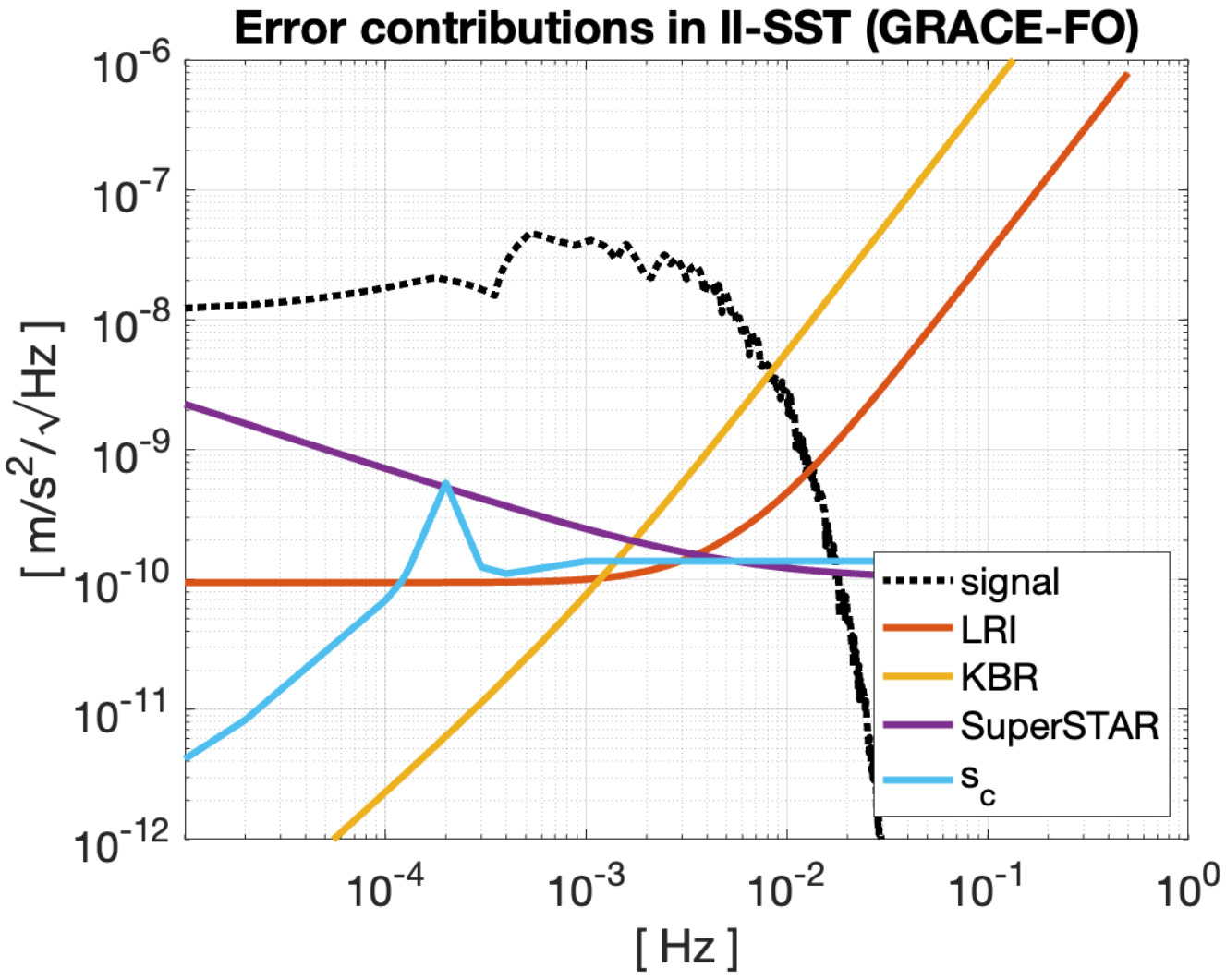

Figure 11: Error amplitude spectra for GRACE-FO's LRI (red line), KBR (yellow line), ranging instruments, the SuperSTAR accelerometer (purple line), and the error in reconstructing the centrifugal accelerations (green line), compared to the estimated time-variable signal magnitude (dotted black line).

The errors in reconstructing the centrifugal accelerations (light blue line) are not dominant over the whole frequency bands, which is expected because this error type is not reported in literature as detrimental to the quality of the estimated gravity field models. On the other hand, it should be noted that the component related to the accuracy of the radial velocity $s_{v_z}$ is, by far, the dominant term in Eq. ( 26 ). Consequently, $s_c$ is intrinsic to the ll-SST concept that uses the leader/follower formation (also known as *trailing* or *inline*) and has the reported magnitude irrespective of the ISR and accelerometer accuracies.

### 3.1.2. ll-SST with quantum accelerometers

The ll-SST concept is limited by the performance of the electrostatic accelerometers at longer wavelengths (frequencies near and below the orbital period), which motivates replacing them with quantum accelerometers to eliminate this limitation (Nicklaus et al., 2019). Contrary to the electrostatic accelerometers, the quantum ones do not need to be calibrated. Therefore, it is sufficient to place one quantum accelerometer into the satellite CoM, as shown in Figure 12, which has the benefit that the accelerometer directly measures the non-gravitational acceleration and is insensitive to centrifugal and Euler forces and gravity gradients. However, depending on the operational mode (Section 2.3.2.1), the atom cloud may be moving during the interferometric measurement process, and we need to account for the effects of the Coriolis force (Section 3.1.2.2).

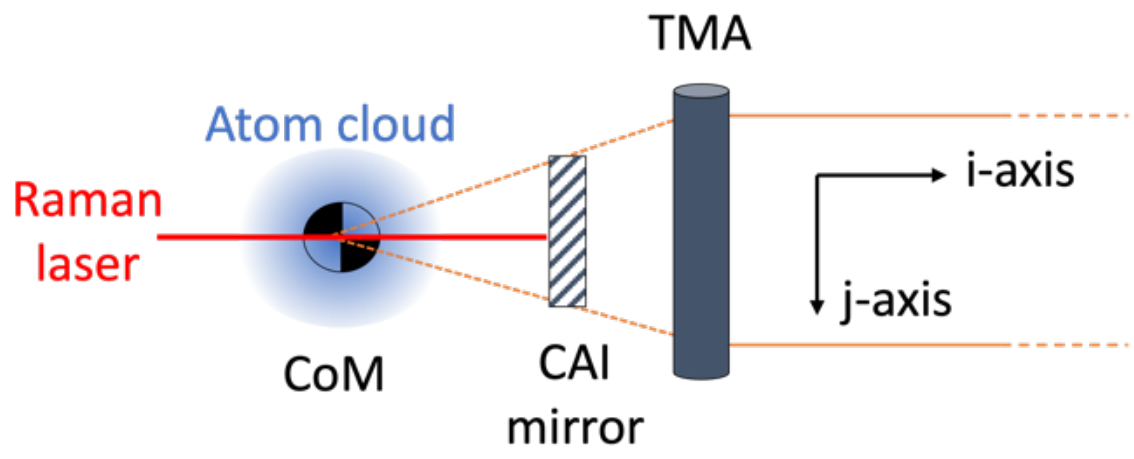

Figure 12: Diagram of the assumed concept of ll-SST with quantum accelerometers. Only one satellite is shown.

The quantum accelerometer measures the phase $\Phi$ of the CAI, which is proportional to the acceleration of the atom cloud $\boldsymbol{a}_{\mathrm{cloud}}$ relative to the acceleration of the mirror that reflects the laser $\boldsymbol{a}_{\mathrm{mirror}}$, the momentum (vector) transfer $\boldsymbol{dp}$ and the square of interrogation time $T$ between laser pulses:

$$\Phi = \frac{\boldsymbol{dp}}{\hbar} \cdot (\boldsymbol{a}_{\mathrm{cloud}} - \boldsymbol{a}_{\mathrm{mirror}})T^2, \tag{27}$$

where $\boldsymbol{dp}$ is the momentum (vector) transfer that defines the direction in which the acceleration is sensed, and its magnitude is given by Equation ( 18 ).
Since the mirror is firmly attached to the satellite, unlike the atom cloud, it serves as a reference for the non-gravitational accelerations $\boldsymbol{a}_{\mathrm{ng}}$ experienced by the satellite so that:

$$\boldsymbol{a}_{\mathrm{mirror}} = \boldsymbol{a}_{\mathrm{ng}}. \tag{28}$$

In contrast, the atom cloud freely floated in inertial space during the interrogation time. The acceleration of the atom cloud can be expressed as:

$$\boldsymbol{a}_{\mathrm{cloud}} = -\left(\boldsymbol{V} - \boldsymbol{\Omega}^2 - \dot{\boldsymbol{\Omega}}\right)(\boldsymbol{r}_{\mathrm{cloud}} - \boldsymbol{r}_{\mathrm{CoM}}) + 2\boldsymbol{\omega} \times \boldsymbol{v}_{\mathrm{cloud}}, \tag{29}$$

with the gravity gradient tensor as:

$$\boldsymbol{V} = \begin{bmatrix} V_{ii} & V_{ij} & V_{ik} \\ V_{ij} & V_{jj} & V_{jk} \\ V_{ik} & V_{jk} & V_{kk} \end{bmatrix}, \tag{30}$$

the angular velocity vector as:

$$\boldsymbol{\omega} = \begin{bmatrix} \omega_i \\ \omega_j \\ \omega_k \end{bmatrix},$$

the angular rate tensor as:

$$\boldsymbol{\Omega} = \begin{bmatrix} 0 & -\omega_k & \omega_j \\ \omega_k & 0 & -\omega_i \\ -\omega_j & \omega_i & 0 \end{bmatrix} \text{ and } \boldsymbol{\Omega}^2 = \begin{bmatrix} -\omega_j^2 - \omega_k^2 & \omega_i\omega_j & \omega_i\omega_k \\ \omega_i\omega_j & -\omega_k^2 - \omega_i^2 & \omega_j\omega_k \\ \omega_i\omega_k & \omega_j\omega_k & -\omega_i^2 - \omega_j^2 \end{bmatrix} \tag{31}$$

and the angular acceleration as:

$$\dot{\boldsymbol{\Omega}} = \begin{bmatrix} 0 & -\dot{\omega}_k & \dot{\omega}_j \\ \dot{\omega}_k & 0 & -\dot{\omega}_i \\ -\dot{\omega}_j & \dot{\omega}_i & 0 \end{bmatrix}. \tag{32}$$

The remaining symbols are the position of the atom cloud $\boldsymbol{r}_{\mathrm{cloud}}$, position of the satellite CoM $\boldsymbol{r}_{\mathrm{CoM}}$, and the velocity of the atom cloud $\boldsymbol{v}_{\mathrm{cloud}}$ relative to the satellite CoM. The term $\dot{\boldsymbol{\Omega}}$ represents the angular acceleration matrix which causes Euler accelerations, $\boldsymbol{\Omega}^2$ causes the centrifugal accelerations, and $2\boldsymbol{\omega} \times \boldsymbol{v}_{\mathrm{cloud}}$ is the Coriolis acceleration.
Combining the above, we obtain the measured acceleration along the direction defined by the Raman laser axis, represented by the unit vector $\boldsymbol{e}_i = \boldsymbol{dp}/dp$, as:

$$\boldsymbol{e}_i \cdot \boldsymbol{a}_{\mathrm{ng}} = \\ = -\frac{\hbar}{dpT^2}\Phi + \boldsymbol{e}_i \cdot \left(-\left(V - \boldsymbol{\Omega}^2 - \dot{\boldsymbol{\Omega}}\right)(\boldsymbol{r}_{\mathrm{cloud}} - \boldsymbol{r}_{\mathrm{CoM}}) + 2\boldsymbol{\omega} \times \boldsymbol{v}_{\mathrm{cloud}}\right). \tag{33}$$

The derivation of the equation above can be found in Section 8.1.1.
Since we measure the acceleration in the direction of the momentum transfer $\boldsymbol{dp}$, this vector must be aligned with the laser used for measuring the ISR. Further, we ignore any effects of magnetic fields and self-gravity on the atom cloud, which might play a role in view of the extreme sensitivity of the quantum sensor. These considerations are beyond the scope of this study because they heavily depend on the specific instrument and satellite design.
If the atom cloud is in the satellite CoM, i.e., $\boldsymbol{r}_{\mathrm{cloud}} - \boldsymbol{r}_{\mathrm{CoM}} = 0$, the equation above simplifies to:

$$\boldsymbol{e}_i \cdot \boldsymbol{a}_{\mathrm{ng}} = -\frac{\hbar}{dpT^2}\Phi + \boldsymbol{e}_i \cdot (2\boldsymbol{\omega} \times \boldsymbol{v}_{\mathrm{cloud}}), \tag{34}$$

leaving the Coriolis term as the only effect to consider. In this context, we note that $\boldsymbol{r}_{\mathrm{cloud}} - \boldsymbol{r}_{\mathrm{CoM}} = 0$ holds for the initial atom cloud position. The first laser pulse of the CAI splits the wave-packet into two that move at a similar speed in opposite directions along the laser axis away from the initial position, indicated as step ii in Section 2.3.2. We assume that the rate of rotation of the Raman laser, i.e., the rotation of the satellite after the compensation by the tilting mirror, does not change significantly during the outward and inward wave-packet drift, indicated as steps ii to iv in Section 2.3.2. Under these conditions, the integrated effect of the rotational and gravity gradient terms cancel out over the outward and inward motion of the two wave-packets and are negligible at their recombination.
Considering axis i, j, and k are arbitrary, e.g., $i, j, k \equiv x, y, z$ for along-track ll-SST and $i, j, k \equiv y, z, x$ for cross-track ll-SST, we can express the equation in scalar form as:

$$a_{\mathrm{ng},i} = -\frac{\hbar}{dpT^2}\Phi + 2\omega_j v_{\mathrm{cloud},k} - 2\omega_k v_{\mathrm{cloud},j}. \tag{35}$$

#### 3.1.2.1. Error amplitude of quantum accelerometers

Considering the error in the atom cloud velocity knowledge $\sigma_{v,\mathrm{cloud}}$, angular rate error $\sigma_\omega$ after compensation by the tilting mirror, and applying error propagation (and considering the effect of the discrete CAI measurement cycle period $T_{\mathrm{cycle}}$), we obtain:

$$s^2_{a_{\mathrm{ng},i}} = \frac{T_{\mathrm{cycle}}}{T^4}\frac{s^2_\Phi}{\beta^2 k^2_{\mathrm{eff}}} + \\ +4\omega_j^2 s^2_{v,\mathrm{cloud},k} + 4\omega_k^2 s^2_{v,\mathrm{cloud},j} + 4s^2_{\omega,j} v^2_{\mathrm{cloud},k} + 4s^2_{\omega,k} v^2_{\mathrm{cloud},j}. \tag{36}$$

We can group the first term as the CAI acceleration sensitivity error $\sigma_{\mathrm{CAI,ng}}$ (cf. Section 2.3.2) and the last four terms as the errors $s_{\mathrm{Cor},i}$ caused by the Coriolis effect:

$$s^2_{a_{\mathrm{ng},i}} \equiv s^2_{\mathrm{CAI,ng},i} + s^2_{\mathrm{Cor},i}. \tag{37}$$

#### 3.1.2.2. Amplitude of the Coriolis term

The equation of the variance of the Coriolis term depends on the velocity of the atom cloud and the angular velocity of the Raman laser, cf. Equation ( 36 ). In this section, we analyse the individual contributions in more detail and explain the underlying assumptions on signals and errors.

#### 3.1.2.2.1. Cloud velocity

The atom cloud velocity components that contribute to the variance of the Coriolis term are perpendicular to the Raman laser axis, since we assume that the integrated effects from the outwards and inwards movement of the cloud along the Raman laser axis are negligible.
Since no instrument is perfect, the atom cloud has a random, non-zero velocity when released from the MOT. We assume the worst-case value of the initial cloud velocity to be $s_{v_{\text{cloud,initial}}} = 10^{-7}\text{m/s}$. It should be noted that treating this worst-case value as an error is a conservative estimate if a technique called Delta Kick-Collimation (DKC) is used (Amri, 2022).
The thermal velocity (most probable speed) of individual atoms is:

$$v_{\text{atom, therm}} = \sqrt{\frac{2k_B T_{\text{atom}}}{m}}, \quad (38)$$

where $k_B$ is the Stefan Boltzmann constant, $T_{\text{atom}}$ is the atom temperature, and $m$ is the atom mass. In practice, $v_{\text{atom,therm}}$ tends to be at the order of $10^{-4}$ m/s, but DKCallows for values at the micrometre per second (Amri, 2022). For this reason, we assume that $v_{\text{atom,therm}} = 10^{-6}$ m/s.
Recalling that the velocity dispersion follows the Maxwell-Boltzmann distribution (Amri, 2022), we are interested in the difference between the most probable speed $v_{\text{atom,therm}}$ and the velocity Root Mean Square (RMS), which is given by $\sqrt{3/2}v_{\text{atom,therm}}$. Under this assumption, the thermal velocity of the atom cloud is:

$$s_{v_{\text{cloud,therm}}} = \frac{\left(\sqrt{\frac{3}{2}} - 1\right) v_{\text{atom, therm}}}{\sqrt{N}}, \quad (39)$$

due to the averaging over $N$ atoms. Thus, assuming that $s_{v_{\text{cloud,initial}}}$ is uncorrelated to $s_{v_{\text{cloud,therm}}}$, the variability of the atom cloud velocity is:

$$s_{v_{\text{cloud}}}^2 = s_{v_{\text{cloud,therm}}}^2 + s_{v_{\text{cloud,initial}}}^2. \quad (40)$$

#### 3.1.2.2.2. Angular velocity

We remind that the angular velocity $\omega_i$ is not necessarily the satellite angular velocity but the residual angular velocity after compensation by the tilting mirror if the mission concepts under analysis assume this capability. To make this distinction, we introduce the symbol $\delta\omega$ when relevant, but derive all equations with the original symbol for angular velocity $\omega$.
The mean angular velocity of a nadir-pointing satellite in LEO is the pitch rate of about $\omega_y = 1.1 \times 10^{-3}\text{rad/s}$, with the $y$-axis aligned with the cross-track direction. The yaw and roll rate are typically at least one order of magnitude smaller, i.e., about $\omega_x = \omega_z = 1 \times 10^{-4}\text{rad/s}$ (Allasio et al., 2010), with the $x$-axis aligned with the along-track direction and the $z$-axis right-hand orthogonal (which is parallel to the radial direction in a circular orbit).
The compensation of the angular velocity requires that either the satellite attitude or the tilting mirrors are controlled, taking input from a sensor that provides angular velocity measurements. One of the best-performing angular velocity sensors is the Astrix 200 laser gyroscope, which has an accuracy of $s_\omega = 5 \times 10^{-8}\text{rad/s}$. We believe that it will not be possible to fully exploit the gyroscope's performance in real-time and assume an accuracy degradation given by the factor $\text{f}_\omega$:

$$\delta\omega = \text{f}_\omega s_\omega, \quad (41)$$

taken to be one order of magnitude, $\text{f}_\omega = 10$, worse than the measurement system, i.e., $\delta\omega = 5 \times 10^{-7}\text{rad/s}$ for the case of the mirror compensating the satellite rotation driven by the Astrix 200 gyroscope. It is a conservative assumption since experiments in the China Space

Station have shown comparative error levels between the attitude derived from the piezoelectric mirror and an external gyroscope (Li et al., 2025). When the satellite rotation is compensated in this way, we assume $\boldsymbol{\omega} = \delta\boldsymbol{\omega}$. In addition to avoiding loss of contrast in the CAI, these figures already suggest that the compensation of the satellite rotation is necessary for quantum accelerometers, depending on the sensitivity to the Coriolis effect, to be quantified in Section 5.1.

We only consider attitude compensation for the cases of quantum accelerometry or gradiometry and not for electrostatic accelerometers or classical gradiometers, for example, by physically rotating the instrument within the satellite body.

## 3.2. Gradiometry

One important distinction between classical and short-term quantum gradiometry is that in the former case, the accelerometers are sensitive to all directions. Exceptions exist (Barrett et al., 2019), although their technical implementation is exponentially more complex. This results in the possibility of measuring off-diagonal gravity gradient components, even with a single gradiometer arm. There are theoretical studies on vector quantum accelerometers (Barrett et al., 2019), but we assume this technology will be unavailable in the near future. In the case of quantum gravimetry, the CAI instruments we consider are inherently unidirectional and only sensitive to the gravity gradient along the axis connecting the atom clouds. Three CAI gradiometer instruments installed perpendicular to each other in the same satellite are needed to retrieve all diagonal terms of the gravity gradient tensor. For simplicity, we assume this to be the case without going into more detail regarding the engineering aspects of this configuration. For more information on the configuration, the reader is invited to read Trimeche et al. (2019).

### 3.2.1. Gradiometry with electrostatic accelerometers

The GOCE mission demonstrated GG based on electrostatic accelerometers. The configuration of accelerometers is illustrated in Figure 13. Here, we assume the same concept but with more advanced accelerometers and, potentially, a high-performance gyroscope. For reference, we take a distance of $L_{\mathrm{GG}} = 0.5$ m between the accelerometers located on the same axis.

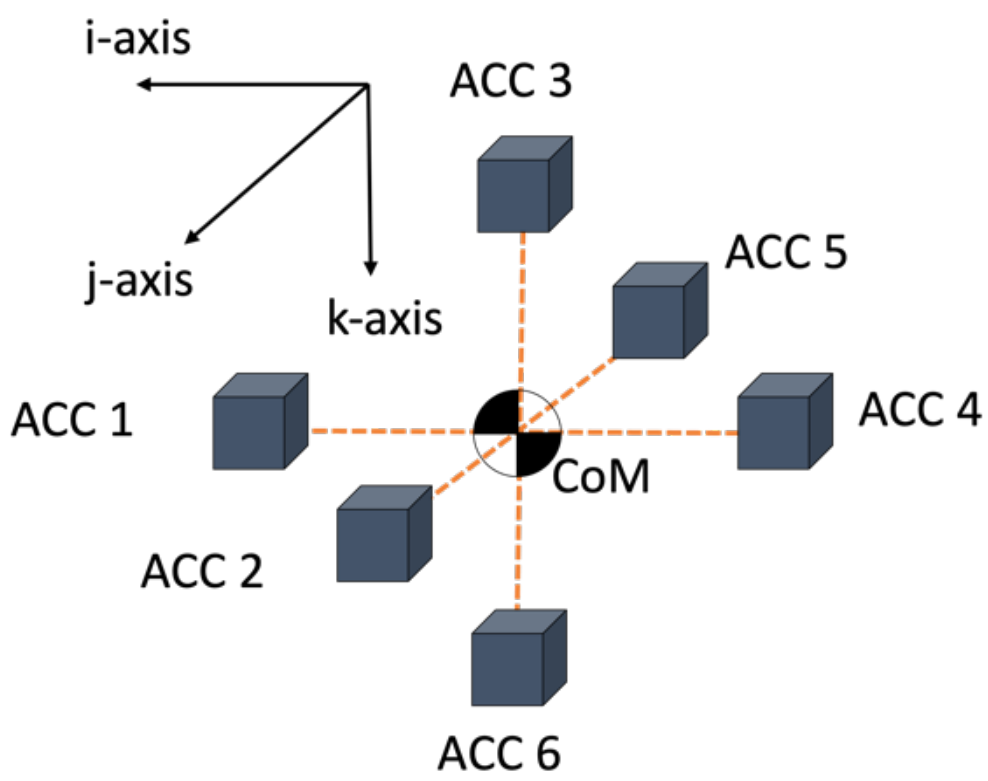

Figure 13: Accelerometer configuration in the classical GG concept.

#### 3.2.1.1. Error amplitude of classical gradiometers

The diagonal element $V_{ii}$ of the gravity gradient tensor, measured by the accelerometer pair 1 and 4, is calculated with (Rummel et al., 2011):

$$V_{ii} = -\frac{a_{1,i} - a_{4,i}}{L_{\mathrm{GG}}} - \omega_j^2 - \omega_k^2. \quad (42)$$

For the remaining diagonal elements $V_{jj}$ and $V_{kk}$, the expressions are similar to accelerometers pairs 2,5 and 3,6, respectively, and angular velocity components *i,k* and *i,j*, respectively. After error propagation, there are non-linear terms that result in the product of angular velocity and angular velocity noise components, following approximations such as:

$$(\omega_i + s_{\omega,i})^2 \approx \omega_i^2 + 2\omega_i s_{\omega,i}. \tag{43}$$

This relation illustrates the coupling between the angular velocity signal $\omega_i$ and the angular velocity noise $\sigma_{\omega,i}$, along the generic axis $i$.
The errors associated with classical gradiometry are:

$$s_{V_{ii}}^2(f) = \frac{2}{L_{\text{GG}}^2} s_{\text{acc,ng}}^2(f) + 4\left(\left(\text{RMS}(\omega_j)\right)^2 + \left(\text{RMS}(\omega_k)\right)^2\right) s_{\omega}^2(f), \tag{44}$$

where $\text{RMS}(\omega_j)$ and $\text{RMS}(\omega_k)$ are the RMS of the angular velocity signal, which signifies the coupling between signal and noise in the non-linear error propagation. For the pitch rate, we assume $\text{RMS}(\omega_y) \approx 1.1 \times 10^{-3}\ rad/s$ and for the yaw and roll rate, $\text{RMS}(\omega_x) \approx \text{RMS}(\omega_z) \approx 10^{-4}\ rad/s$. The underlying assumptions for deriving these RMS values are explained in Appendix 8.1. For the remaining diagonal terms, the corresponding expressions are similar to Equation ( 44 ), considering the RMS of the orthogonal angular velocity components.
The off-diagonal elements of the gravity gradient tensor are given by:

$$\begin{aligned} V_{ij} &= -\frac{a_{2,i} - a_{5,i}}{2L_{\text{GG}}} - \frac{a_{1,j} - a_{4,j}}{2L_{\text{GG}}} + \omega_i\omega_j \\ V_{ik} &= -\frac{a_{1,k} - a_{4,k}}{2L_{\text{GG}}} - \frac{a_{3,i} - a_{6,i}}{2L_{\text{GG}}} + \omega_i\omega_k \\ V_{jk} &= -\frac{a_{3,j} - a_{6,j}}{2L_{\text{GG}}} - \frac{a_{2,k} - a_{5,k}}{2L_{\text{GG}}} + \omega_j\omega_k \end{aligned} \tag{45}$$

Applying error propagation results in:

$$s_{V_{ij}}^2(f) = \frac{1}{L_{GG}^2} s_{\text{acc,ng}}^2(f) + \left(\left(\text{RMS}(\omega_i)\right)^2 + \left(\text{RMS}(\omega_j)\right)^2\right) s_{\omega}^2(f). \tag{46}$$

The expressions for the remaining off-diagonal components are similar, with the RMS of the angular velocity components $\omega_i$, $\omega_k$ and $\omega_j$, $\omega_k$ relevant to the tensor components *ik* and *jk*, respectively.

#### 3.2.1.2. Attitude reconstruction with classical gradiometers

We analyse the capability of the classical gradiometer to measure angular acceleration and separate gravity gradients from frame rotations. It is fair to assume that the satellite is equipped with star sensors for attitude reconstruction, both on board and in-ground processing. We presume the errors of this instrument are given by Equation ( 3 ).
The GG also functions by design as an accurate angular acceleration sensor, from which the angular velocity can be obtained by numerical integration. The noise ASD of the gradiometer-derived angular accelerations is derived from the accelerometer measurements as follows:

$$\begin{aligned} \dot{\omega}_i &= \frac{a_{2,k} - a_{5,k}}{2L_{\text{GG}}} - \frac{a_{3,j} - a_{6,j}}{2L_{\text{GG}}}, \\ \dot{\omega}_j &= \frac{a_{3,i} - a_{6,i}}{2L_{\text{GG}}} - \frac{a_{1,k} - a_{4,k}}{2L_{\text{GG}}}, \\ \dot{\omega}_k &= \frac{a_{1,j} - a_{4j}}{2L_{\text{GG}}} - \frac{a_{2,i} - a_{5,i}}{2L_{\text{GG}}}. \end{aligned} \tag{47}$$

Since all accelerometer axes have the same performance, we find:

$$s_{\dot{\omega},GG}(f) = \frac{1}{L_{\text{GG}}} s_{\text{acc,ng}}(f), \tag{48}$$

for the noise ASD of the gradiometer-derived angular acceleration measurements, where $s_{\text{acc,ng}}(f)$ is the noise ASD of a linear acceleration measurement of an individual accelerometer. It is worthwhile to note that the angular acceleration measurements derived in this way are more accurate than the angular acceleration measurements of the individual accelerometers, as illustrated in Figure 14, taking MicroSTAR (Section 2.3.1.2) as an example.

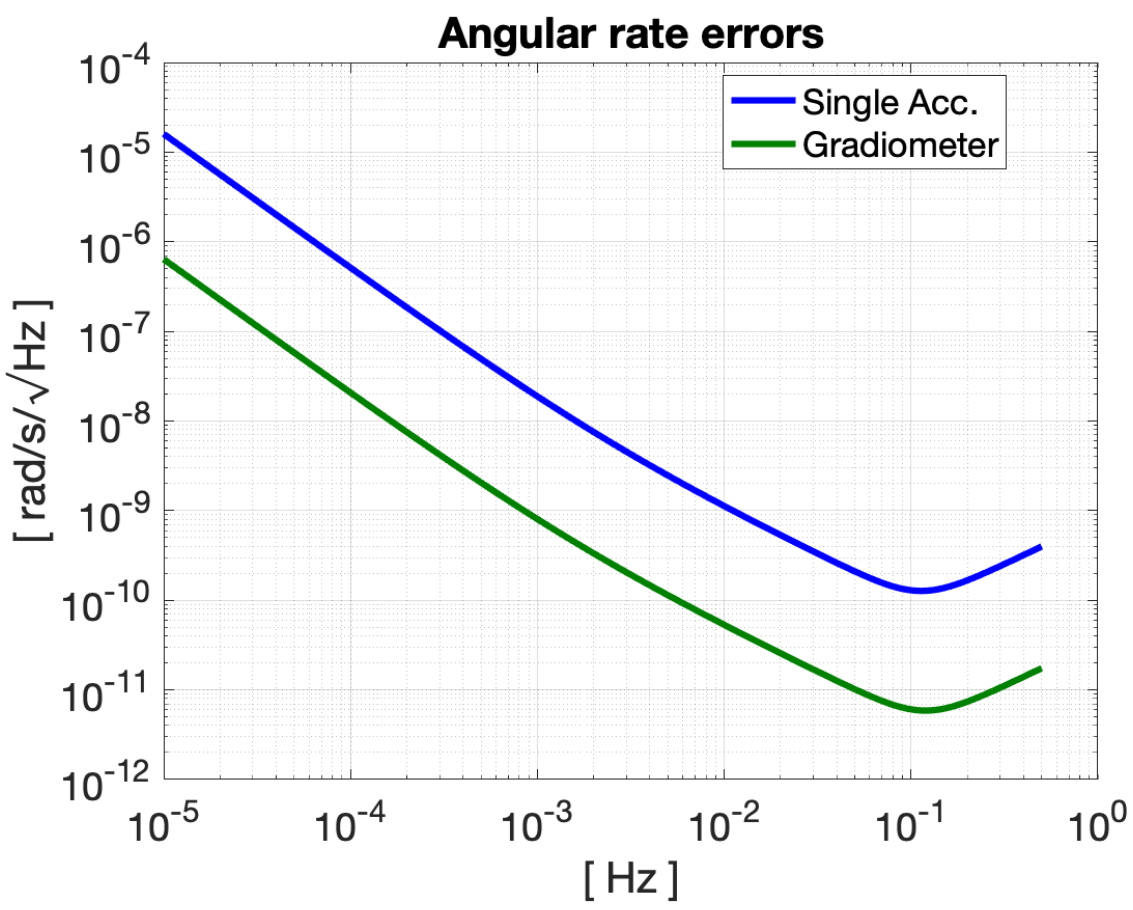

Figure 14: Noise spectra of the angular velocity derived from the linear acceleration measurements of two accelerometers installed in a gradiometer arm with length $L_{\text{GG}} = 0.5$ m (green line) and of the angular acceleration measurements of a single accelerometer (blue line).

The discrepancy between the attitude accuracies in Figure 14 can be explained by the small distance (a few cm) between the electrodes for the single accelerometer and the larger gradiometer arm (50 cm) separating the accelerometers.

### 3.2.2. Gradiometry with Quantum Sensors

GG based on quantum sensors relies on generating an initial cloud of atoms that is physically split into two atom clouds, labelled "Cloud 1" and "Cloud 2" in Figure 15. The advantage is that the relative cloud positions are well-known because the splitting is performed by a laser pulse that gives an accurately known kick to the initial cloud of atoms. Once the two atom clouds are in position, the cold atom interferometric sequence (cf. Section 2.3.2) starts for both atom clouds simultaneously using the same Raman laser. In principle, it is the same sequence of laser pulses as for quantum accelerometers. Thus, the quantum gradiometer is also based on sensing the differential acceleration along the laser axis over the precisely known distance between the atom clouds. Consequently, the equations presented in Section 2.3.2 also apply to quantum GG.

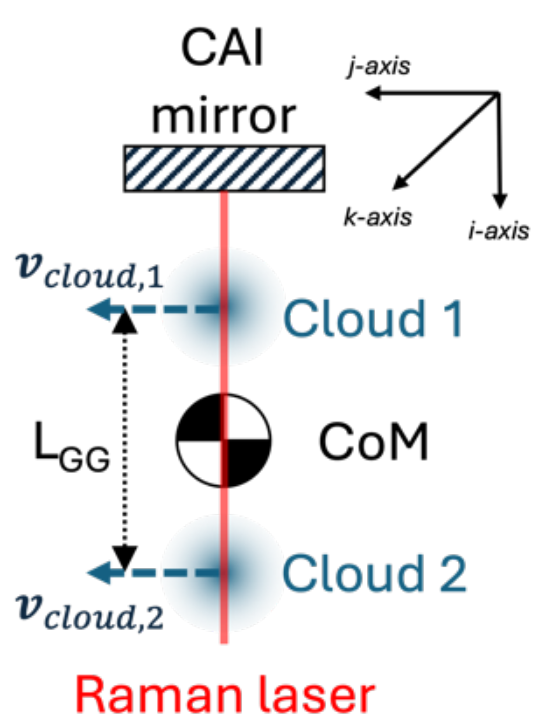

Figure 15: Quantum GG concept.

By rearranging Equation ( 33 ), we find:

$$\frac{\hbar}{dpT^2}\Phi_{i,l} = \boldsymbol{e}_i \cdot \left(-\boldsymbol{a}_{\mathrm{ng,l}} - \left(V - \boldsymbol{\Omega}^2 - \dot{\boldsymbol{\Omega}}\right)\left(\boldsymbol{r}_{\mathrm{cloud,l}} - \boldsymbol{r}_{\mathrm{CoM}}\right) + 2\boldsymbol{\omega} \times \boldsymbol{v}_{\mathrm{cloud,l}}\right), \qquad (49)$$

where $l = 1$ or 2 identifies the atom cloud (cf. Figure 15). Setting the origin of the reference axis at the CoM, the atom clouds are at positions such that $\boldsymbol{r}_{\mathrm{cloud,1}} - \boldsymbol{r}_{\mathrm{cloud,2}} \equiv L_{\mathrm{GG}}/2\boldsymbol{e}_i$ at the time the interferometry is done.

Regarding the cloud velocity, we have $\boldsymbol{v}_{\mathrm{cloud,1}} = \boldsymbol{v}_{\mathrm{cloud,2}}$ and:

$$\boldsymbol{v}_{\mathrm{cloud},l} = \boldsymbol{v}_{\mathrm{cloud,therm},l} + \boldsymbol{v}_{\mathrm{cloud,initial},l} = \boldsymbol{v}_{\mathrm{cloud,therm},l} + v_{\mathrm{cloud,initial},l} \cdot \boldsymbol{e}_j. \qquad (50)$$

In the case of the concurrent mode of operation (Section 2.3.2.1), $\boldsymbol{v}_{\mathrm{cloud,initial},l}$ reflects the transverse velocity imparted by an additional laser perpendicular to the Raman laser axis. In this case, we note that $v_{\mathrm{cloud,initial,1}} = v_{\mathrm{cloud,initial,2}}$ because both clouds are accelerated with the same recoil laser (not shown in Figure 15). In the case of the sequential mode, $v_{\mathrm{cloud,initial,1}} = v_{\mathrm{cloud,initial,2}} = 0$. In both cases, $\boldsymbol{a}_{\mathrm{ng,1}} = \boldsymbol{a}_{\mathrm{ng,2}}$ because the non-gravitational accelerations affect both clouds equally. As a result, these two terms cancel when computing the differential measurement $\delta\Phi_i \equiv \Phi_{i,1} - \Phi_{i,2}$:

$$\frac{\hbar}{dpT^2}\delta\Phi_i = \boldsymbol{e}_i \cdot \left(-\left(\boldsymbol{V} - \boldsymbol{\Omega}^2 - \dot{\boldsymbol{\Omega}}\right)L_{\mathrm{GG}}\boldsymbol{e}_i + 2\boldsymbol{\omega} \times \boldsymbol{v}_{\mathrm{cloud,therm}}\right), \qquad (51)$$

The term on $\boldsymbol{v}_{\mathrm{cloud,therm},l}$ does not cancel because it is related to the thermal velocity of the cloud, which is akin to a random variable.

Recognising that $\dot{\boldsymbol{\Omega}}\boldsymbol{e}_i = \dot{\boldsymbol{\Omega}}\boldsymbol{e}_j = \dot{\boldsymbol{\Omega}}\boldsymbol{e}_k = 0$ (cf. Equation ( 32 )), isolating the gravity gradient term, and reducing the vector equation to a scalar quantity results in:

$$V_{ii} = -\frac{1}{L_{\mathrm{GG}}}\frac{\hbar}{dpT^2}\delta\Phi_i - \omega_j^2 - \omega_k^2 + \frac{2}{L_{\mathrm{GG}}}\left(\omega_j v_{\mathrm{cloud,therm},k} - \omega_k v_{\mathrm{cloud,therm},j}\right). \qquad (52)$$

The derivation of the equation above and the one for other axes is in Appendix 8.1.2. As a side note, if we assume a quantum GG concept where the satellite rotation is compensated by a tilting mirror, as described in Section 3.1.2, the symbol $\omega$ should be interpreted as the compensated attitude $\delta\omega$, given by Equation ( 41 ).

#### 3.2.2.1. Error amplitude of quantum gradiometers

Applying error propagation to Equation ( 51 ), under the assumption that Raman laser is rotating at the residual angular velocity after compensation by the tilting mirror $\delta\boldsymbol{\omega}$, and considering uncorrelated phase errors, $s_{\delta\Phi_i}^2 = s_{\Phi_{i,1}}^2 + s_{\Phi_{i,2}}^2 = 2s_{\Phi}^2$, results in the errors associated with quantum gradiometry:

$$\begin{aligned} s_{V_{ii}}^2 = {} & \frac{2}{L_{\mathrm{GG}}^2}\frac{T_{\mathrm{cycle}}}{T^4}\frac{s_{\Phi_i}^2}{\beta^2 k_{\mathrm{eff}}^2} + 4\omega_j^2 s_{\omega_j}^2 + 4\omega_k^2 s_{\omega_k}^2 \\ & + \frac{16}{L_{\mathrm{GG}}^2}(\omega_j^2 s_{v_{\mathrm{cloud,therm}},k}^2 + \omega_k^2 s_{v_{\mathrm{cloud,therm}},j}^2 + \\ & s_{\omega,j}^2 v_{\mathrm{cloud,therm},k}^2 + s_{\omega,k}^2 v_{\mathrm{cloud,therm},j}^2). \end{aligned} \qquad (53)$$

Under the conservative assumption of homogeneous noise in all components:

$$\begin{aligned} s_V^2 & = \frac{2}{L_{\mathrm{GG}}^2}\frac{T_{\mathrm{cycle}}}{T^4}\frac{s_{\Phi_i}^2}{\beta^2 k_{\mathrm{eff}}^2} + 8\omega^2 s_\omega^2 + \frac{32}{L_{\mathrm{GG}}^2}\left(\omega^2 s_{v_{\mathrm{cloud,therm}}}^2 + s_\omega^2 v_{\mathrm{cloud,therm}}^2\right) \\ & \equiv s_{V_{\mathrm{CAI}}}^2 + s_{\Omega^2}^2 + s_{V_{\mathrm{Cor}}}^2 = \frac{2}{L_{\mathrm{GG}}^2}s_{\mathrm{CAI,ng}}^2 + s_{\Omega^2}^2 + \frac{4}{L_{\mathrm{GG}}^2}s_{\mathrm{Cor}}^2, \end{aligned} \qquad (54)$$

with expressions for $s_{\mathrm{CAI,ng}}$ and $s_{\mathrm{Cor}}^2$ given in Section 3.1.2.1, and the errors associated with the centrifugal accelerations are $s_{\Omega^2} = \sqrt{8}\omega s_\omega$.

#### 3.2.2.2. Attitude reconstruction with quantum gradiometers

In the case of classical gradiometry (Section 3.2.1), the 6 capacitive accelerometers that compose the 3D gradiometer make it possible to estimate the rate of change of the attitude (Section 3.2.1.2). For CAI gradiometry, the case is not the same since the "accelerometers" are unidimensional. This section explores how single-axis quantum gradiometers can be arranged so that they also provide complete attitude information.

Consider the $m$-th CAI gradiometer aligned with the $i$-axis (parallel to $\boldsymbol{e}_i$), shown in Figure 16.

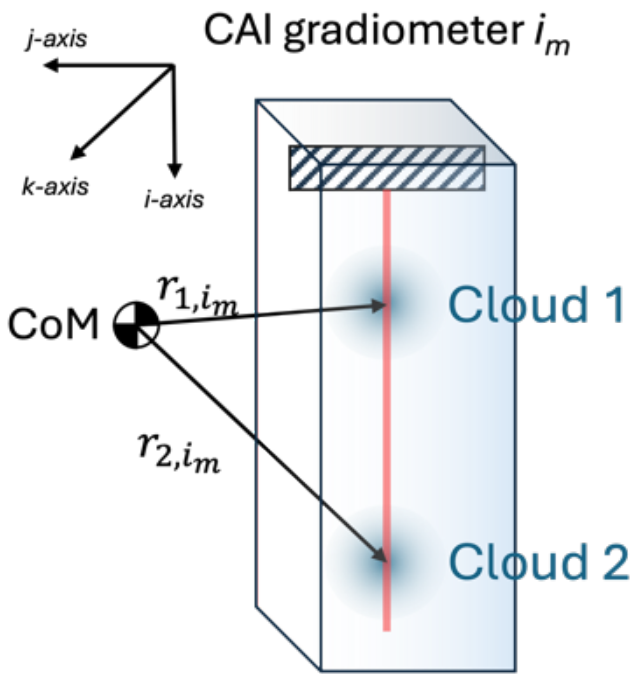

Figure 16: CAI gradiometer $m$, aligned with axis $i$, showing the positions of the atom clouds 1 and 2.

From Equation ( 29 ), with the reference frame at the CoM, the phase measurement $\Phi_{l,i_m}$ of the atom cloud $l$ is:

$$\frac{\Phi_{l,i_m}}{k_{\mathrm{eff}}T^2} = \boldsymbol{e}_i \cdot \left(-\left(\boldsymbol{V} - \boldsymbol{\Omega}^2 - \dot{\boldsymbol{\Omega}}\right)\boldsymbol{r}_{l,i_m} + 2\boldsymbol{\omega} \times \boldsymbol{v}_{\mathrm{cloud}}\right) =$$
$$\boldsymbol{e}_i \cdot \begin{bmatrix} -V_{ii} - \omega_j^2 - \omega_k^2 & -V_{ij} + \omega_i\omega_j - \dot{\omega}_k & -V_{ik} + \omega_i\omega_k + \dot{\omega}_j \\ -V_{ij} + \omega_i\omega_j + \dot{\omega}_k & -V_{jj} - \omega_i^2 - \omega_k^2 & -V_{jk} + \omega_j\omega_k - \dot{\omega}_i \\ -V_{ik} + \omega_i\omega_k - \dot{\omega}_j & -V_{jk} + \omega_j\omega_k + \dot{\omega}_i & -V_{kk} - \omega_i^2 - \omega_j^2 \end{bmatrix} \boldsymbol{r}_{l,i_m} + \qquad (55)$$
$$\boldsymbol{e}_i \cdot \begin{bmatrix} \omega_j v_{l,i_m,k} - \omega_k v_{l,i_m,j} \\ \omega_k v_{l,i_m,i} - \omega_i v_{l,i_m,k} \\ \omega_i v_{l,i_m,j} - \omega_j v_{l,i_m,i} \end{bmatrix} + \boldsymbol{e}_i \cdot \boldsymbol{a}_{\mathrm{ng}}.$$

Any linear combination of the two phase measurements $\Phi_{1,i_m}$ and $\Phi_{2,i_m}$ unavoidably includes off-diagonal terms of $\boldsymbol{V}$, $\boldsymbol{\Omega}^2$ and $\dot{\boldsymbol{\Omega}}$ because $\boldsymbol{r}_{l,i_m}$ is not aligned with $\boldsymbol{e}_i$. Therefore, there are 15 unknowns:

- 6 gravity gradients $\boldsymbol{V}$,
- 3 angular rates $\boldsymbol{\omega}$ (or accelerations $\dot{\boldsymbol{\Omega}}$),
- 3 non-gravitational accelerations $\boldsymbol{a}_{\mathrm{ng}}$ and
- 3 Coriolis accelerations $\boldsymbol{\omega} \times \boldsymbol{v}_{\mathrm{cloud}}$ (or atom cloud velocities $\boldsymbol{v}_{\mathrm{cloud}}$).

In the case of the common and differential linear combinations, 3 degrees of freedom cancel out: the diagonal of the gravity gradient and the non-gravitational accelerations, respectively. In both cases, we are left with 12 unknowns, which require 12 CAI gradiometers to resolve fully.

One possible arrangement of CAI gradiometers is indicated in Figure 17 as a "ring" of 4 gradiometer arms arranged symmetrically around the CoM in the same plane, perpendicular to the $k$-axis and intersecting the CoM.

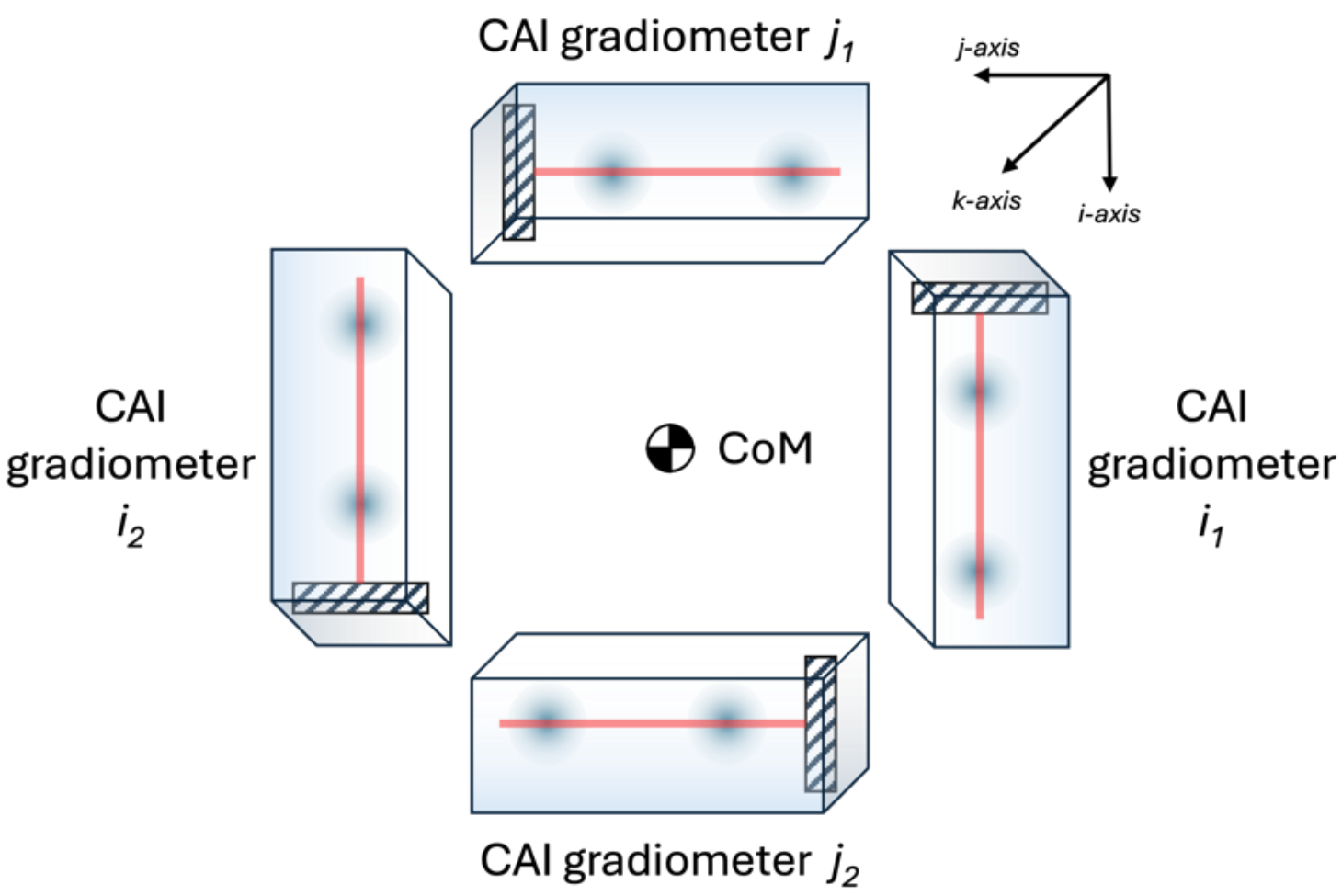

Figure 17: Arrangement of 4 CAI gradiometers placed symmetrically around the CoM that are able to measure $a_{\mathrm{ng},i}$ , $a_{\mathrm{ng},j}$ and $\dot{\omega}_k$.

The common phase measurement $\sum \Phi_{i1}$ (or $\sum \Phi_{i2}$) and $\sum \Phi_{j1}$ (or $\sum \Phi_{j2}$) of gradiometer $i_1$ (or $i_2$) and $j_1$ (or $j_2$) provides information on the $i$-th and $j$-th component of the non-gravitational accelerations, respectively. The difference between the common phases $\sum \Phi_{i1} - \sum \Phi_{i2}$ or $\sum \Phi_{j1} - \sum \Phi_{j2}$ cancels the Coriolis and non-gravitational accelerations. The sum of the difference of the common phases $\sum \Phi_{i1} - \sum \Phi_{i2} + \sum \Phi_{j1} - \sum \Phi_{j2}$ cancels the gravity gradients and centrifugal accelerations, isolating the $\dot{\omega}_k$ term and, consequently, $\omega_k$ by numerical integration. It should be noted that the complete attitude of the satellite could only be determined with the help of an absolute attitude sensor, such as a star tracker, to resolve the attitude bias and bias rate in the case of the proposed CAI gradiometer ring, resulting from the unknown integration constant.
If two orthogonal rings measure the $k$ and $j$ components of the angular acceleration, that would be sufficient to completely retrieve the measurement of the $V_{ii}$, which would ideally be aligned with the radial direction. The reason is because $\omega_i$ does not affect the measurement of $V_{ii}$. If a third orthogonal ring is added, then the complete diagonal of $\boldsymbol{V}$ is available.

# 4. Effects of attitude uncertainty

So far, we have derived the noise spectra for the observations in the satellite reference frame (SRF). However, the observations need to be transformed to the Earth-centred, Earth-fixed (ECEF) reference frame. The transformation requires satellite attitude data and is, therefore, not free of errors. In this section, we analyse the impact of attitude errors.
The effects of attitude uncertainty discussed in this section are conceptually separate from other attitude errors discussed so far, notably the Coriolis effect on CAI accelerometers or gradiometers (Section 3.1.2.2) and the capability of gradiometers to observe attitude rates, analysed in Section 3.2.1.2 for the classical case, and in Section 3.2.2.2 for a collection of CAI gradiometers.

## 4.1. ll-SST

Conceptually, we need to subtract the ISR acceleration $\ddot{\rho}_{ng}$ due to non-gravitational accelerations from the measured ISR acceleration $\ddot{\rho}$ to obtain $\ddot{\rho}_{\mathrm{grav}}$ due to gravitational accelerations:

$$\ddot{\rho}_{\mathrm{grav}} = \ddot{\rho} + \frac{1}{\rho} v_z^2 - \ddot{\rho}_{\mathrm{ng}}. \qquad (56)$$

For this analysis, we isolate the term $\ddot{\rho}_{\text{ng}}$, i.e., the differential non-gravitational acceleration projected onto the LOS connecting the two satellites' centres of mass, i.e.:

$$\ddot{\rho}_{\text{ng}} = \left(\boldsymbol{a}_{\text{ng}}^{(1)} - \boldsymbol{a}_{\text{ng}}^{(2)}\right) \cdot \boldsymbol{e}_{\text{LOS}} = \Delta\boldsymbol{a}_{\text{ng}} \cdot \boldsymbol{e}_{\text{LOS}}, \qquad (57)$$

where $\boldsymbol{e}_{LOS}$ is the LOS unit vector and $\boldsymbol{a}_{\text{ng}}^{(1)}$ and $\boldsymbol{a}_{ng}^{(2)}$ are the measured non-gravitational accelerations of satellite 1 and satellite 2, respectively, all defined in the Earth-centred, inertial (ECI) reference. The LOS vector is defined by the satellite positions, which are naturally provided in the ECI frame.

$$\boldsymbol{e}_{\text{LOS}} = \frac{\boldsymbol{r}^{(1)} - \boldsymbol{r}^{(2)}}{\|\boldsymbol{r}^{(1)} - \boldsymbol{r}^{(2)}\|}. \qquad (58)$$

If $\boldsymbol{r}^{(1)}$ or $\boldsymbol{r}^{(2)}$ are provided in the ECEF frame, we assume that the coordinate transformations do not introduce a significant error because the Earth Orientation Parameters are well known. As for the non-gravitational accelerations, they are measured in the SRF, which we represent $\boldsymbol{a}_{\text{ng}}^{(\text{SRF,l})}$ for satellite *l*.

We need to perform a coordinate transformation for the accelerations, based on the satellite attitude data, from the SRF to the ECI frame, which can be represented as

$$\boldsymbol{a}_{\text{ng}}^{(\text{ECI,l})} = \boldsymbol{R}^{(\text{ECI}\leftarrow\text{SRF,l})}\boldsymbol{a}_{\text{ng}}^{(\text{SRF,l})}. \qquad (59)$$

The SRF-to-ECI rotation matrix contains errors, here represented by the small angle rotation matrix $\boldsymbol{\Theta}^{(\text{ECI}\leftarrow\text{SRF,s})}$:

$$\boldsymbol{R}^{(\text{ECI}\leftarrow\text{SRF,l})} = \boldsymbol{R}^{(\text{ECI}\leftarrow\text{SRF,l,true})}\boldsymbol{\Theta}^{(\text{ECI}\leftarrow\text{SRF,l})}. \qquad (60)$$

We split $\boldsymbol{\Theta}^{(\text{ECI}\leftarrow\text{SRF,l})}$ into the identity matrix and the small angle rotations, given as Euler angles errors $\theta_{\text{roll,l}}$, $\theta_{\text{pitch,l}}$ and $\theta_{\text{yaw,l}}$ for satellite *l*:

$$\boldsymbol{\Theta}^{(\text{ECI}\leftarrow\text{SRF,l})} = \boldsymbol{E}^{(l)} + \boldsymbol{I} = \begin{bmatrix} 0 & -\theta_{\text{yaw,l}} & \theta_{\text{pitch,l}} \\ \theta_{\text{yaw,l}} & 0 & -\theta_{\text{roll,l}} \\ -\theta_{\text{pitch,l}} & \theta_{\text{roll,l}} & 0 \end{bmatrix} + \boldsymbol{I}. \qquad (61)$$

With these definitions, we can write:

$$\boldsymbol{a}_{\text{ng}}^{(\text{ECI},l)} = \boldsymbol{a}_{\text{ng}}^{(\text{ECI,true},l)} + \boldsymbol{E}^{(l)}\boldsymbol{a}_{\text{ng}}^{(\text{SRF},l)}, \qquad (62)$$

where the only term on the right-hand side that is not perfectly known (for this analysis) is $\boldsymbol{E}^{(l)}\boldsymbol{a}_{\text{ng}}^{(\text{SRF},l)}$. Replacing in Equation( 57 ):

$$\ddot{\rho}_{\text{ng}}^{(ECI)} = \left(\Delta\boldsymbol{a}_{\text{ng}}^{(\text{ECI,true})} + \left(\boldsymbol{E}^{(1)} - \boldsymbol{E}^{(2)}\right)\Delta\boldsymbol{a}_{\text{ng}}^{(\text{SRF})}\right) \cdot \boldsymbol{e}_{\text{LOS}}^{(\text{ECI})}, \qquad (63)$$

Assuming the attitude errors of $\boldsymbol{E}^{(l)}$, represented by the vector $\boldsymbol{s}_{\theta}$, are the same for both satellites, the error propagation of Equation ( 63 ) is

$$s_{\ddot{\rho}_{\text{ng}},\theta}^{(ECI)} = \boldsymbol{\nabla}\ddot{\rho}_{\text{ng}}^{(ECI)}[s_{\theta}]\boldsymbol{\nabla}\ddot{\rho}_{\text{ng}}^{(ECI)^T}, \qquad (64)$$

with the Jacobian $\boldsymbol{\nabla} = \left[\frac{\partial}{\partial\theta_{\text{roll}}} \; \frac{\partial}{\partial\theta_{\text{pitch}}} \; \frac{\partial}{\partial\theta_{\text{yaw}}}\right]$ and $[s_{\theta}]$ the diagonal matrix with the errors of $\theta_{\text{roll}}$, $\theta_{\text{pitch}}$ and $\theta_{\text{yaw}}$, i.e., $s_{\theta_{\text{roll}}}$, $s_{\theta_{\text{pitch}}}$, and $s_{\theta_{\text{yaw}}}$.

Evaluating Equation ( 64 ), we arrive at the ranging error due to attitude $s_{\ddot{\rho}_{\text{ng}},\theta}$ as a function of $\boldsymbol{s}_{\theta} = \left[s_{\theta_{\text{roll}}} \; s_{\theta_{\text{pitch}}} \; s_{\theta_{\text{yaw}}}\right]^{\text{T}}$:

$$s_{\ddot{\rho}_{\text{ng}},\theta}^{(ECI)} = \sqrt{2}\begin{bmatrix} \Delta a_{\text{ng,y}}^{(\text{SRF})} e_{\text{LOS,z}}^{(\text{ECI})} - \Delta a_{\text{ng,z}}^{(\text{SRF})} e_{\text{LOS,y}}^{(\text{ECI})} \\ \Delta a_{\text{ng,x}}^{(\text{SRF})} e_{\text{LOS,z}}^{(\text{ECI})} - \Delta a_{\text{ng,z}}^{(\text{SRF})} e_{\text{LOS,x}}^{(\text{ECI})} \\ \Delta a_{\text{ng,x}}^{(\text{SRF})} e_{\text{LOS,y}}^{(\text{ECI})} - \Delta a_{\text{ng,y}}^{(\text{SRF})} e_{\text{LOS,x}}^{(\text{ECI})} \end{bmatrix}^{\text{T}} \boldsymbol{s}_{\theta}. \quad (65)$$

We make the conservative assumption that the amplitude of $\Delta\boldsymbol{a}_{\text{ng}}^{(\text{SRF})}$ is given by the RMS of the non-gravitation accelerations at orbital altitude and take the component with the largest magnitude in each entry of the row vector in Equation ( 65 ). We further assume that the amplitude of $\boldsymbol{e}_{\text{LOS}}^{(\text{ECI})}$ is 1 and drop the superscript of the reference frame because the errors have the same amplitude in any frame. Equation ( 65 ) simplifies to

$$\rho_{\ddot{\rho}_{\text{ng}},\theta} = \sqrt{2}\begin{bmatrix} \max\left(\text{RMS}(a_{\text{ng,z}}),\text{RMS}(a_{\text{ng,z}})\right) \\ \max\left(\text{RMS}(a_{\text{ng,x}}),\text{RMS}(a_{\text{ng,z}})\right) \\ \max\left(\text{RMS}(a_{\text{ng,x}}),\text{RMS}(a_{\text{ng,y}})\right) \end{bmatrix}^{\text{T}} \boldsymbol{s}_{\theta}. \quad (66)$$

In the case of homogenous attitude error, $s_{\theta_{\text{yaw}}} = s_{\theta_{\text{pitch}}} = s_{\theta_{\text{roll}}} \equiv s_{\theta}$, for example, if there are multiple star tracker cameras and their data is combined optimally with additional attitude sensors, such as an IMU and DWS:

$$s_{\ddot{\rho}_{\text{ng}},\theta} = \sqrt{2}\text{RMS}(a_{\text{ng,x}}) s_{\theta}. \quad (67)$$

The values for $\text{RMS}(a_{\text{ng,x}})$, $\text{RMS}(a_{\text{ng,y}})$ and $\text{RMS}(a_{\text{ng,z}})$ we considered in this study are presented in Appendix 8.3 and are functions of the Drag-Free Control (DFC) system. We quantify the effect of these errors in Section 5.2 for both electrostatic and quantum accelerometry.

## 4.2. Gradiometry

In the case of GG, the gravity gradients observed in the SRF $\boldsymbol{V}^{(\text{SRF})}$ are related to the gravity gradients in the ECEF reference frame $\boldsymbol{V}^{(\text{ECEF})}$ by:

$$\boldsymbol{V}^{(\text{ECEF})} = \boldsymbol{R}^{(\text{ECEF}\leftarrow\text{SRF})}\boldsymbol{V}^{(\text{SRF})}\boldsymbol{R}^{(\text{ECEF}\leftarrow\text{SRF})^{T}}, \quad (68)$$

where $\boldsymbol{R}^{(\text{ECEF}\leftarrow\text{SRF})}$ is the rotation that transforms from the satellite to the ECEF reference frame. The attitude measurements generally relate the ECI frame to the SRF frame, so it makes more sense to split the rotation in these frames:

$$\boldsymbol{V}^{(\text{ECEF})} = \boldsymbol{R}^{(\text{ECEF}\leftarrow\text{ECI})}\boldsymbol{R}^{(\text{ECI}\leftarrow\text{SRF})}\boldsymbol{V}^{(\text{SRF})}\boldsymbol{R}^{(\text{ECI}\leftarrow\text{SRF})^{T}}\boldsymbol{R}^{(\text{ECEF}\leftarrow\text{ECI})^{T}}. \quad (69)$$

Assuming the ECI to ECEF frame is known perfectly, we can focus on the gravity gradients in the ECI frame:

$$\boldsymbol{V}^{(\text{ECI})} = \boldsymbol{R}^{(\text{ECI}\leftarrow\text{SRF})}\boldsymbol{V}^{(\text{SRF})}\boldsymbol{R}^{(\text{ECI}\leftarrow\text{SRF})^{T}}. \quad (70)$$

As for ll-SST, Section 4.1, we model the transformation $\boldsymbol{R}^{(\text{ECI}\leftarrow\text{SRF})}$ as the product of the error-free rotation $\boldsymbol{R}^{(\text{ECI}\leftarrow\text{SRF,true})}$ and a small-angle rotation matrix $\boldsymbol{\Theta}^{(\text{ECI}\leftarrow\text{SRF})}$:

$$\boldsymbol{R}^{(\text{ECI}\leftarrow\text{SRF})} = \boldsymbol{R}^{(\text{ECI}\leftarrow\text{SRF,true})}\boldsymbol{\Theta}^{(\text{ECI}\leftarrow\text{SRF})}, \quad (71)$$

and Equation ( 68 ) becomes:

$$\boldsymbol{V}^{(\text{ECI})} = \boldsymbol{R}^{(\text{ECI}\leftarrow\text{SRF,true})}\boldsymbol{\Theta}^{(\text{ECI}\leftarrow\text{SRF})}\boldsymbol{V}^{(\text{SRF})}\boldsymbol{\Theta}^{(\text{ECI}\leftarrow\text{SRF})^{\text{T}}}\boldsymbol{R}^{(\text{ECI}\leftarrow\text{SRF,true})^{\text{T}}}. \quad (72)$$

If we restrict our analysis to the terms that contain errors, we can safely ignore the error-free transformation $\boldsymbol{R}^{(\text{ECI}\leftarrow\text{SRF,true})}$:

$$\boldsymbol{V}^{(\text{ECI})} = \boldsymbol{\Theta}^{(\text{ECI}\leftarrow\text{SRF})}\boldsymbol{V}^{(\text{SRF})}\boldsymbol{\Theta}^{(\text{ECI}\leftarrow\text{SRF})^{\text{T}}}, \quad (73)$$

We split $\boldsymbol{\Theta}^{(\mathrm{ECI}\leftarrow\mathrm{SRF})} = \boldsymbol{E} + \boldsymbol{I}$ as before and drop the reference frame superscript:

$$\boldsymbol{V}^{(\mathrm{ECI})} = (\boldsymbol{I} + \boldsymbol{E})\, \boldsymbol{V}^{(\mathrm{SRF})} (\boldsymbol{I} + \boldsymbol{E})^{\mathrm{T}} \tag{74}$$

As usual, the error propagation of Equation ( 74 ) requires the tensor $\boldsymbol{V}^{(\mathrm{ECI})}$ to be collapsed into the vector $\boldsymbol{v}^{(\mathrm{ECI})}$, producing the error $9 \times 9$ covariance matrix $\boldsymbol{C}_{\boldsymbol{V}}$:

$$\boldsymbol{C}_{\boldsymbol{V}} = \nabla \boldsymbol{v}^{(\mathrm{ECI})} [s_{\theta}] \nabla {\boldsymbol{v}^{(\mathrm{ECI})}}^{\mathrm{T}} \tag{75}$$

In evaluating Equation ( 75 ), we assume that $\boldsymbol{V}$ is error-free, the small angles are negligible $\theta_{\mathrm{roll,s}} = \theta_{\mathrm{pitch,s}} = \theta_{\mathrm{yaw,s}} \approx 0$, and ignore the cross-correlations such that $[s_{\boldsymbol{V},\theta}] = \mathrm{diag}(\boldsymbol{C}_{\boldsymbol{V}})$, resulting in

$$\begin{bmatrix} s^2_{V_{xx},\theta} \\ s^2_{V_{yy},\theta} \\ s^2_{V_{zz},\theta} \\ s^2_{V_{xy},\theta} \\ s^2_{V_{xz},\theta} \\ s^2_{V_{yz},\theta} \end{bmatrix} = \begin{bmatrix} 0 & 4V_{xz}^2 & 4V_{xy}^2 \\ 4V_{yz}^2 & 0 & 4V_{xy}^2 \\ 4V_{yz}^2 & 4V_{xz}^2 & 0 \\ V_{xz}^2 & V_{yz}^2 & (V_{xx} - V_{yy})^2 \\ V_{xy}^2 & (V_{xx} - V_{zz})^2 & V_{yz}^2 \\ (V_{yy} - V_{zz})^2 & V_{xy}^2 & V_{xz}^2 \end{bmatrix} \begin{bmatrix} s^2_{\theta_{\mathrm{roll}}} \\ s^2_{\theta_{\mathrm{pitch}}} \\ s^2_{\theta_{\mathrm{yaw}}} \end{bmatrix}. \tag{76}$$

Considering the amplitude of the gravity gradient signal presented in Section 8.4 and setting the errors $\boldsymbol{s}_{\theta}$ equal to 1, the scaling of the attitude errors into gravity gradient errors is:

$$[s_{\boldsymbol{V},\theta}] = \begin{bmatrix} 0.2 & 5 & 3600 \\ & 9.7 & 3603 \\ & & 9.7 \end{bmatrix} \times 10^{-3}\, E$$

# 5. Results and discussion

We present our results by quantifying the amplitude of frame accelerations, i.e., those related to the effect of the Coriolis accelerations and the centrifugal accelerations in Section 5.1. We quantify the errors for ll-SST in Section 5.2 for both electrostatic (Section 5.2.1) and quantum (Section 5.2.2) accelerometers. Finally, we quantify the errors for quantum gradiometry in Section 5.3 for the gradiometer operating in sequential mode.

## 5.1. Importance of the frame accelerations

In this section, we quantify the amplitude of the Coriolis and centrifugal accelerations. CAI accelerometry is only affected by the former (Section 3.1.2.1), while gradiometry is affected by both (Section 3.2.2.1).

Recall that the variance of the Coriolis term $\sigma^2_{\mathrm{Cor},i}$ in a CAI accelerometer aligned with the *i*-axis, cf. Equations ( 36 ) and ( 37 ), are:

$$s^2_{\mathrm{Cor},i} = 4\omega_j^2 s^2_{v,\mathrm{cloud},k} + 4\omega_k^2 s^2_{v,\mathrm{cloud},j} + 4 s^2_{\omega,j} v^2_{\mathrm{cloud},k} + 4 s^2_{\omega,k} v^2_{\mathrm{cloud},j} \tag{77}$$

which is valid for both along-track and cross-track ll-SST.

In Section 3.1.2.2.2, we discussed the variances $s^2_{\omega,j}$ and $s^2_{\omega,k}$ reflect the angular velocity measurement noise, for which we assume $s_{\omega,j} = s_{\omega,k} = 5 \times 10^{-8}\mathrm{rad/s}$ in the case of using the high-performance Astrix 200 laser gyroscope, here assumed to be white noise for simplicity. For the magnitude of the angular velocity after tilting mirror compensation $\delta\omega$, we proposed one order of magnitude worse performance than the errors, i.e., $\delta\omega = 5 \times 10^{-7}\mathrm{rad/s}$, in the *full attitude compensation* scenario. In the *no tilting mirror* scenario, we assume $\omega = 1 \times 10^{-4}\mathrm{rad/s}$ for yaw and roll and $\omega = 1.1 \times 10^{-3}\mathrm{rad/s}$ for pitch (cf. Section 3.1.2.2.2). We also consider the intermediate case of *minimum pitch-rate compensation*, where the pitch is compensated to the level of $\delta\omega = 1 \times 10^{-4}\mathrm{rad/s}$.

For the atom cloud velocity, we assume that either one component is $v_{\text{cloud,j}} = 2.5$ cm/s and the other components are zero in the case of the concurrent operational mode, or all atom cloud velocity components are zero in the case of the sequential mode of operation (cf. Section 2.3.2.1). For the uncertainty of the atom cloud velocity $s_{v_{\text{cloud}}}$, we assume DKC with $v_{\text{atom,therm}} = 10^{-6}$ m/s and $s_{v_{\text{cloud,initial}}} = 10^{-7}$m/s (cf. Section 3.1.2.2.1, Equations ( 39 ) and ( 40 )), resulting in $s_{v_{\text{cloud,therm}}} = 2.3 \times 10^{-10}$ m/s for $N = 10^6$.
Under these assumptions, we can quantify the effect of the Coriolis term for the concurrent and sequential operational modes combined with different levels of attitude compensation, as summarised in Table 1.

Table 1: Standard deviation of the Coriolis term $\sigma_{\text{Cor},i}$, assuming $\sigma_{\omega,j} = \sigma_{\omega,j} = 5 \times 10^{-8}$ rad/s, $v_{\text{atom,therm}} = 10^{-6}$ m/s and $\sigma_{v_{\text{cloud,initial}}} = 10^{-7}$m/s for several combinations of angular velocity compensation scenarios and operational modes (affecting the cloud velocity), for the case of along-track ll-SST and the *i*-axis aligned with the along-track direction.

| Attitude compensation scenario | Residual angular velocity [rad/s] | Concurrent mode [m/s²] | Sequential mode [m/s²] |
|---|---|---|---|
| | | $v_{\text{cloud,k}} = \sigma_{v_{\text{cloud,therm}}} = 0.23$nm/s $v_{\text{cloud,j}} = 2.5$cm/s | $v_{\text{cloud,k}} = v_{\text{cloud,j}} = \sigma_{v_{\text{cloud,therm}}} = 0.23$nm/s |
| No tilting mirror | $\omega_j = 1.1 \times 10^{-3}$ $\omega_k = 10^{-4}$ | $2.5 \times 10^{-9}$ | $2.2 \times 10^{-10}$ |
| Minimum pitch-rate compensation | $\delta\omega_j = \omega_k = 10^{-4}$ | $2.5 \times 10^{-9}$ | $2.8 \times 10^{-11}$ |
| Full attitude compensation | $\delta\omega_j = \delta\omega_k = = 5 \times 10^{-7}$ | $2.5 \times 10^{-9}$ | $2.0 \times 10^{-13}$ |

In the concurrent case, the Coriolis effect is dominated by the large cloud velocity and is insensitive to attitude compensation. Therefore, the only possibility for a CAI accelerometer to outperform the MicroSTAR accelerometer, which has a precision of $2 \times 10^{-12}$m/s$^2$ (cf. Section 2.3.1.2), is to consider full attitude compensation and zero atom cloud velocity provided by the sequential mode of operation. This choice limits the measurement cycle to be equal to the interrogation time, as explained in Section 2.3.2.1.
To make the concurrent mode of operation competitive, one would have to reduce the initial cloud velocity to at least $10^{-5}$ m/s for the Coriolis effects to reduce to the level of $10^{-12}$ m/s$^2$. This extremely slow velocity would increase the sampling time prohibitively; one may as well cycle through interferometry and atom production in a sequential way. The only other option is to decrease the attitude uncertainty by 3 orders of magnitude, which is very technically challenging for classical attitude sensors, e.g., ground equipment is only at best 2 orders of magnitude better (Stedman et al., 2003).
For CAI gradiometry, the effect of the Coriolis accelerations in Equation ( 54 ) is:

$$s_{V_{\text{Cor}}} = \frac{2}{L_{GG}} s_{\text{Cor}}, \qquad (78)$$

which effectively means that noise in the gravity gradients is a factor of 4 worse compared to CAI accelerometers, assuming $L_{GG} = 0.5$m. In addition to that, the problem is exacerbated by the small gravity gradient time-variable signal shown in Figure 9. A gradiometer operating in sequential mode with full attitude compensation would have $s_{V_{\text{Cor}}} = 0.8$ mE $= 8.0 \times 10^{-13}$s$^{-2}$, which is insufficient to sense the time-variable gravity field. We note that this discussion is exclusively based on the effect of the Coriolis force, with no regard to the CAI interferometric sensitivity discussed in Section 2.3.2.

The amplitude of the effect of the centrifugal accelerations is $s_{\Omega^2} = \sqrt{8}\delta\omega\sigma_\omega$, which follows from Equation ( 54 ). Continuing with the assumption that the angular rate has a noise of $s_\omega = 5 \times 10^{-8}\text{rad/s}$ and that $\omega$ is related to the tilting mirror compensation $\delta\omega = 5 \times 10^{-7}\text{rad/s}$, we expect $s_{\Omega^2} = 0.071$ mE. Consequently, unlike the Coriolis forces, the centrifugal accelerations do not limit the CAI gradiometer's sensitivity to temporal variations of the gravity field.

## 5.2. ll-SST

In this analysis, we include the effects of attitude uncertainty presented in Section 4.1, for which the magnitude of non-gravitational accelerations is important. We consider the 2 scenarios motivated in Section 8.3: the RMS of the non-gravitational accelerations experienced at 230 km and the residual non-gravitational accelerations with a 3D DFC system at the same altitude. We note that we have omitted the case of the 1D DFC system, similar to GOCE, because it is identical to the *no DFC* scenario since the y and z-axis non-gravitational accelerations are relevant to the case where the ISR axis is (roughly) aligned with the x-direction. Those are the same for the two considered DFC scenarios, cf. Equation ( 66 ).

### 5.2.1. Electrostatic accelerometry

We start our analysis of the ll-SST future gravimetric missions with the case of electrostatic accelerometry. The three DFC scenarios differently amplify the attitude errors $s_\omega$, for which we assume 3 scenarios: DWS of LISA (Section 2.1.3.1), DWS of GRACE-FO (Section 2.1.3.2), and no DWS. In all scenarios, the measurements of the sensors above are optimally combined with the attitude measurements from the star tracker (Section 2.1.1), Astrix 200 laser gyroscope (Section 2.1.2), and accelerometer (Section 2.1.4). The results are shown in Figure 18, along with the predicted performance of the ISR instrument in 2040 (Section 2.2.4), the errors in reconstructing the centrifugal accelerations (Section 3.1.1.1) and the errors of the MicroSTAR accelerometer (Section 2.3.1.2).

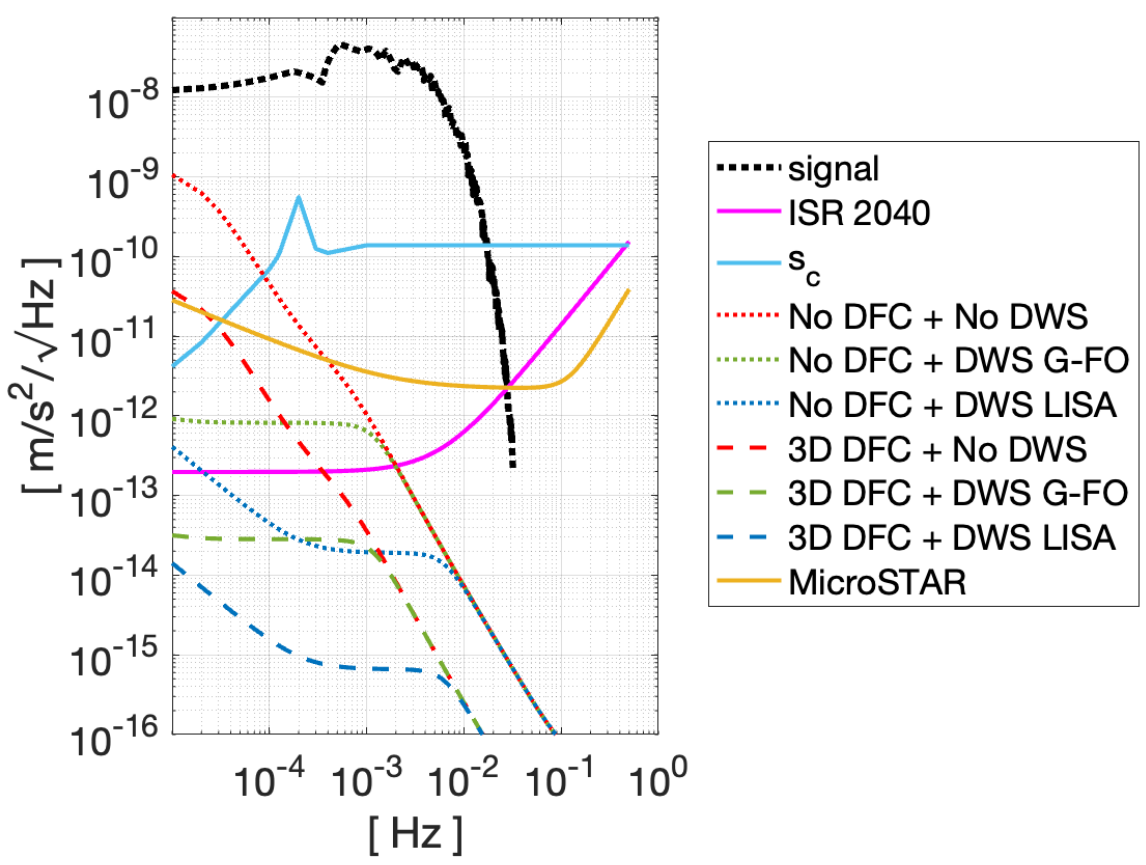

Figure 18: Effect of the attitude errors for no (red, blue and green dotted lines) and 3D (red, blue and green dashed lines) DFC with three different combinations of attitude instruments indicated in the legend plus the attitude derived from the μASC star tracker and accelerometer, compared to the errors of the accelerometer (solid yellow line), ISR errors predicted for 2040 (solid pink line), errors in reconstructing the centrifugal accelerations (light blue line), and the estimated time-variable signal magnitude (dotted black line).

The main message of Figure 18 is that the components of the time-variable gravity field model (black dotted line) are fully observed until 20mHz, or roughly SH degree 113, assuming sufficiently dense ground track coverage in a reasonably short period. At this frequency, the signal represented by the black dotted line crosses both the errors in reconstructing the centrifugal accelerations. We recognise that the model we assumed to represent these errors is extrapolated at these high frequencies and may be overly conservative. Still, we have no evidence to support this hypothesis. We also postulate that, if these errors become dominant, advanced GNSS processing strategies that exploit multi-GNSS and other measurement techniques such as Satellite Laser Ranging (SLR), Very Large Baseline Interferometry (VLBI) and Interferometric Synthetic Aperture Radar (InSAR) and LEO constellations (Ge et al., 2022; Li et al., 2023), may increase the accuracy with which the radial velocity is estimated. If this is the case, the high accuracy of the ISR instrument is not fully utilised below this frequency because of the insufficient accelerometer performance, since the yellow line is above the pink line.

The attitude errors are insignificant for the majority of the scenarios. We predict that the need for DFC is only necessary if DWS is not available (red dotted line), which is unlikely since that has already been demonstrated for GRACE-FO.

### 5.2.2. Quantum accelerometry

In order to make complete use of the high accuracy of the ISR instrument predicted for 2040 (Section 2.2.4), we propose the CAI accelerometer indicated in Table 2.

Table 2: CAI parameters

| Parameter | Equation | Symbol | Value |
|---|---|---|---|
| Laser wavelength | ( 16 ) | $\lambda$ | 780 nm |
| Number of atoms | ( 17 ) | $N$ | $10^7$ |
| Interferometer contrast | ( 17 ) | $C$ | 0.8 |
| Degree of entanglement | ( 17 ) | $\alpha$ | 0.25 |
| Momentum space separation | ( 18 ) | $\beta$ | 2 |
| Interrogation time | ( 20 ) | $T$ | 5 s |
| Measurement cycle period | ( 21 ) | $T_{\text{cycle}}$ | 1 s |
| Atom thermal velocity | ( 38 )( 39 ) | $v_{\text{atom, therm}}$ | $10^{-6}$ m/s |
| Initial cloud velocity error | N/A | $s_{v_{\text{cloud,initial}}}$ | $10^{-7}$m/s |
| Cloud velocity | ( 40 ) | $v_{\text{cloud}}$ | 0 or 2.5 cm/s |
| Attitude accuracy degradation factor | ( 41 ) | $f_\omega$ | 10 |

We selected the parameters in Table 2 so that the noise amplitude of the CAI accelerometer is below the noise floor of the ISR 2040 instrument, as shown in Figure 19. For this analysis, we maintained the DFC and attitude scenarios of Section 5.2.1, with the exception that the accelerometer-derived attitude is not available. For the Coriolis effects, we considered only one scenario: DWS of GRACE-FO (Section 2.1.3.2), star tracker (Section 2.1.1), and Astrix 200 laser gyroscope (Section 2.1.2); the case with LISA DWS yields a reduced amplitude of the Coriolis effects (not shown) but with no change to the interpretation of the results.

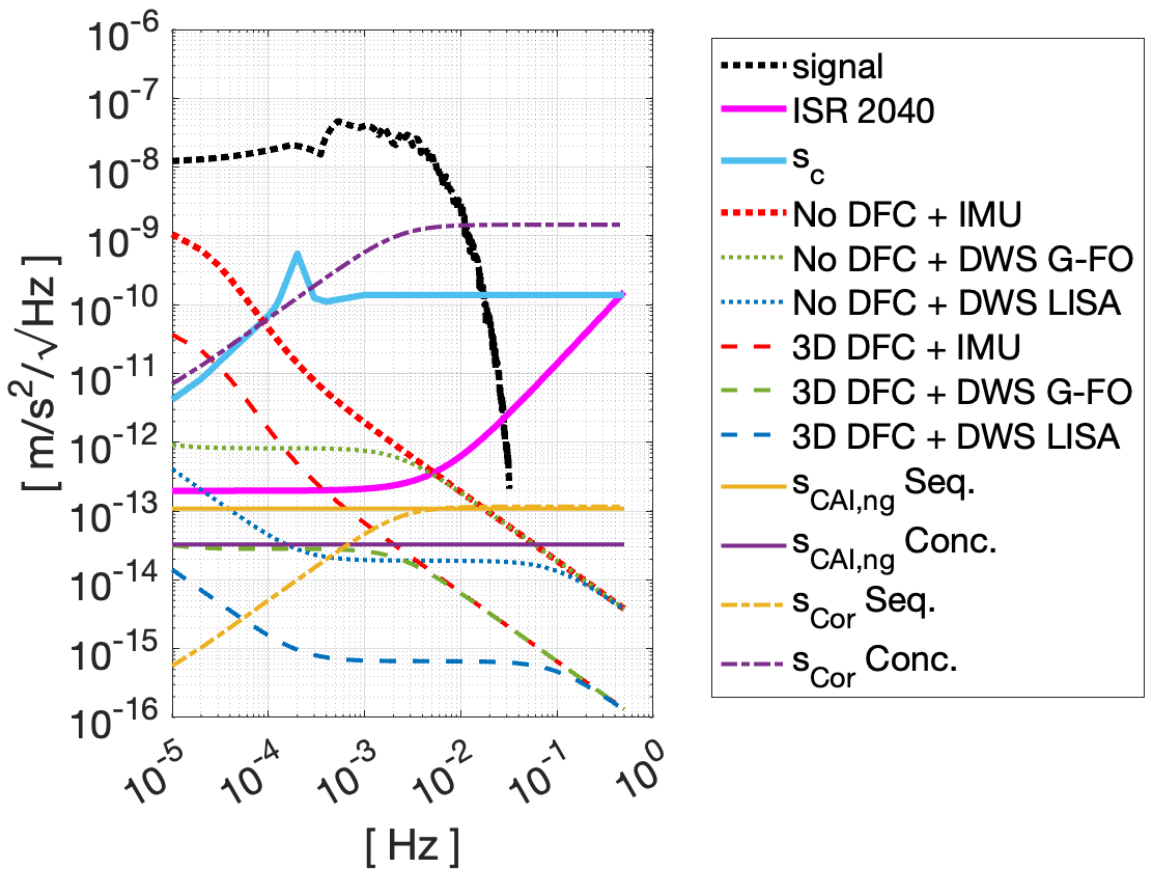

Figure 19: Comparison of the effect of attitude errors (dotted and dashed red, blue and green lines, Section 4.1) with the ISR errors predicted for 2040 (Section2.2.4, solid pink line, identical to Figure 18) and the estimated time-variable signal magnitude (dotted black line, identical to Figure 18), with the errors of the CAI accelerometer operating under the sequential mode (yellow lines) and concurrent mode (purple lines, cf. Section 2.3.2.1), distinguished between the CAI sensitivity (solid purple and yellow lines, Section 2.3.2) and Coriolis effects (dot-dashed purple and yellow lines, Section 3.1.2.2)

In the case of the CAI accelerometer operating in concurrent mode, the noise amplitude is dominated by the Coriolis effect (purple dashed line) due to the non-zero cloud velocity, as explained in Section 5.1, with a noise floor two orders of magnitude above the CAI sensitivity (solid purple line). With such an instrument, not even GRACE's KBR would operate at full capacity, cf. Figure 4. In contrast, this instrument operating in sequential mode has a sensitivity (solid yellow line) a factor of 3 worse than in concurrent mode, as a result of the reduced sampling rate, Equation ( 21 ), but a much-reduced effect of the Coriolis accelerations (dashed yellow line), with a noise floor two orders of magnitude below (4 orders of magnitude if compared to the Coriolis effects of the concurrent mode of operation) and reaching the amplitude of the CAI sensitivity at 3 mHz.

As is the case with the classical ll-SST analysis of Section 5.2.1, the errors in reconstructing the centrifugal accelerations (solid blue line) dominate the error spectrum, except for the Coriolis effects of the CAI accelerometer operating in concurrent mode. We argue that this error may be overestimated at frequencies (much) larger than the orbital period (~ 0.2 mHz) and, if proven necessary, new GNSS processing strategies may likely be able to mitigate it. Neglecting this error at high frequencies, the temporal signal is resolved up to 30 mHz, or SH degree 170, which is the case with the electrostatic case (Section 5.2.1). Unlike the electrostatic case, the quantum accelerometer is more accurate than the LTI at all frequencies, and the time-variable signal is measured with an SNR of $10^4$ at 10 mHz (SH degree 57).

As for attitude and DFC options, the situation is much more demanding than electrostatic accelerometry. For quantum accelerometry, there is a need for LISA-level DWS if DFC is unavailable (dotted blue line). If 3D DFC is available, GRACE-FO's DWS is sufficient (dashed green line). This result illustrates the strict attitude requirements that the increased sensitivity of quantum accelerometers demands on the accuracy of the attitude data.

## 5.3. Gradiometry

In Section 5.1, we quantify the effect of the Coriolis accelerations $s_{V_{\text{Cor}}}$ under the assumption of white noise for the attitude measured by the Astrix 200 laser gyroscope. In reality, the spectra of the errors of this instrument are far from showing constant amplitude with frequency, cf. Section 2.1.2. Additionally, the combination with other attitude instruments was not quantified. We do not consider classical gradiometry because electrostatic accelerometers lack the necessary accuracy to observe the time-variable gravity signal, as shown in Figure 9.
We compare the error spectra of the Coriolis and centrifugal terms with the CAI sensitivity in Table 2, thus making the ll-SST (discussed in Section 5.2.2) and gradiometry cases directly comparable. We only consider the sequential mode of operation (cf. Section 2.3.2.1) because of the destructive effect of the Coriolis accelerations already demonstrated for the ll-SST case. For the impact of attitude uncertainty presented in Section 4.2, we consider the gravity gradient signal amplitudes presented in Section 8.4. Unlike the ll-SST case, drag compensation is not relevant to the errors we analyse in this section because it only affects the non-gravitational signal amplitude. We consider that attitude is measured with the star tracker (Section 2.1.1), Astrix 200 laser gyroscope (Section 2.1.2), and accelerometer (Section 2.1.4). We include the attitude derived from the accelerometer since the demonstration of an early CAI gradiometer in space would benefit from the validation with proven instruments, such as an electrostatic accelerometer. Additionally, this instrument reduces the amplitude of the attitude errors above 0.7 mHz, cf. Figure 1, which is critical for collecting the small time-variable gravity field disturbances. The results of this analysis are presented in Figure 20.

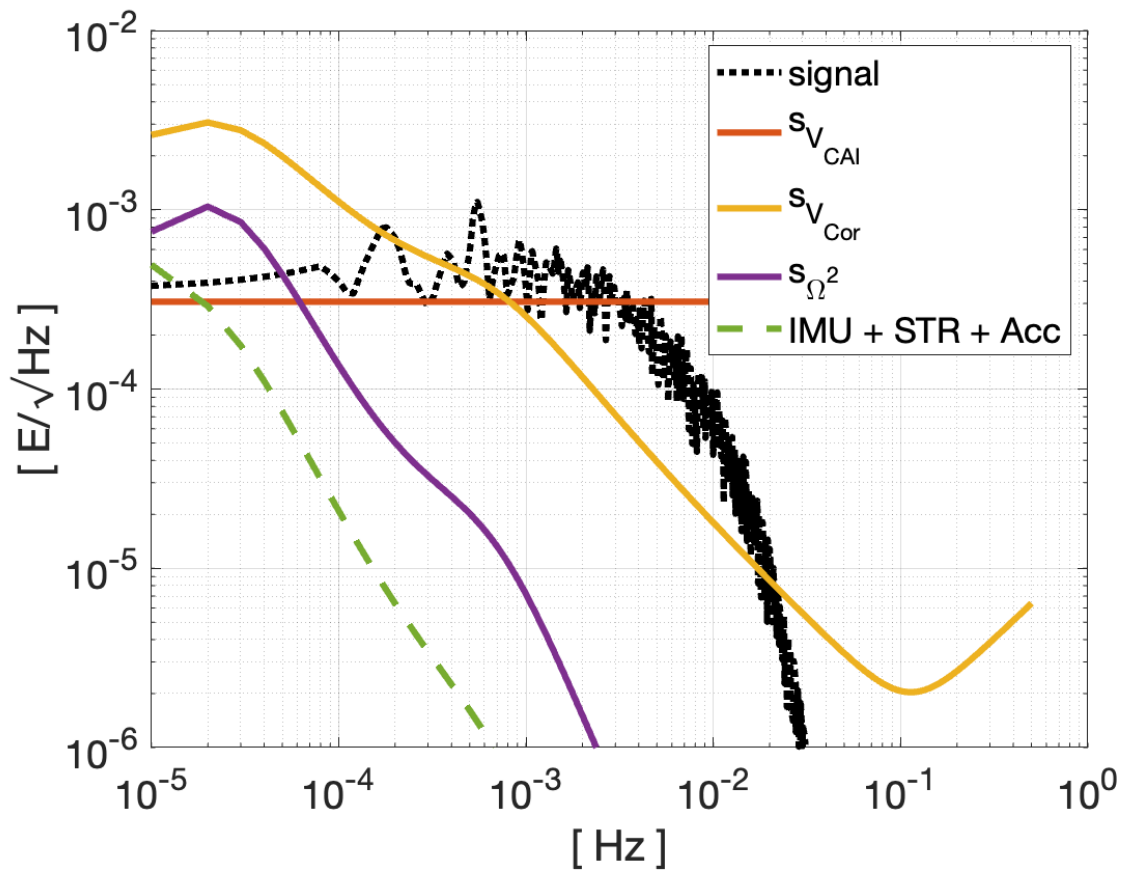

Figure 20: Comparison of the effect of attitude errors (dashed green line, Section 4.2) with the time-variable gravity gradient signal (dotted black line, same as Figure 9) gradiometer CAI sensitivity (red line, Section 3.2.2), the Coriolis effect (yellow line, Section 3.2.2) and the effects of centrifugal accelerations (purple line, Section 3.2.2), as measured by the IMU (Section 2.1.2), star tracker (Section 2.1.1) and attitude derived from the electrostatic accelerometer (Section 2.1.4).

Although the gradiometer CAI sensitivity (red line) is barely enough to resolve the time-variable gravity signal (black dotted line) up to 3 mHz, corresponding roughly to SH degree 17, the Coriolis effects (yellow line) make it impossible to observe this signal below 0.4 mHz or SH degree 2. The impact of attitude uncertainty (purple line) is at least an order of magnitude below the Coriolis effects and only surpasses the magnitude of the gradiometer CAI sensitivity below 0.02 mHz. Of note is that the Coriolis errors (and those associated with centrifugal and attitude errors) are only a function of the attitude sensors and remain the same even if more accurate CAI gradiometers are considered. Therefore, CAI gradiometry requires a novel attitude determination system, which is currently unavailable.
In contrast to the ll-SST case using quantum accelerometers (Section 5.2.2), where the complete signal spectrum is resolved with a high SNR (assuming the the errors in reconstructing the centrifugal accelerations can be properly mitigated), the quantum gradiometer with the same CAI parameters is barely able to determine the time variable signal, with an SNR mostly between 1 and 2, peaking at 3 and dipping at 0.5 at some frequencies.

This example reinforces that the high accuracy of all instruments is critical to the success of CAI gradiometry. Although quantum technology may allow for extremely high CAI sensitivities, a proportional improvement of the attitude sensors is necessary.

# 6. Summary and discussions

In this study, we model the sensitivity of a CAI accelerometer as a function of interferometry contrast, degree of entanglement, number of atoms, momentum space separation and interrogation period (Section 2.3.2). We assume the CAI gradiometer is composed of two CAI accelerometers in the form of two cloud pairs in the same interferometric chamber. In this way, we model quantum gradiometry in a similar way to classical gradiometry, i.e., that the latter is composed of two electrostatic accelerometers. One important difference is that a CAI accelerometer is inherently a one-dimensional instrument, unlike the electrostatic accelerometer. However, this is irrelevant for the diagonal components of the gravity gradient tensor. We intend to make a straightforward application of the currently proposed architectures for CAI instruments, without suggesting or discussing advanced options, such as vectors CAI accelerometers (Barrett et al., 2019), CAI instruments that are also able to observe angular accelerations (Li et al., 2025), or bridging options like classical/quantum hybrid accelerometers (HosseiniArani et al., 2025).

We present electrostatic and quantum accelerometers for ll-SST and gradiometer satellite mission concepts, modelling the measurements and their errors analytically (Section 3). The attitude determination was given special attention, particularly the modelling of the propagation of the associated errors to the observations collected by the accelerometers and gradiometers. For electrostatic accelerometers, the instrument provides attitude information directly since there are multiple electrodes in each facet of the proof mass cavity (Section 2.1.4). On the other hand, so far, no attitude data can be measured by a quantum accelerometer. For classical gradiometry, attitude is estimated with one order of magnitude better accuracy than the electrostatic accelerometer (Section 3.2.1.2). For quantum gradiometry, we demonstrate that 12 uniaxial CAI gradiometers are needed to uniquely resolve the attitude of the satellite (Section 3.2.2.2).

In order to reduce the effects of inaccurately known angular rates on Coriolis accelerations in quantum instruments, defined analytically in Section 3.1.2.1 for CAI accelerometers and in Section 3.2.2.1 for CAI gradiometers, we distinguish between concurrent and sequential modes of operation, with the former allowing for a higher sampling rate and accuracy. The latter minimises atom cloud velocity and, consequently, Coriolis accelerations (Section 2.3.2.1). In Section 5.1, we demonstrate that CAI accelerometry operating in the concurrent mode results in prohibitively high effects resulting from the Coriolis accelerations, limiting the sensitivity of any CAI accelerometer to $s_{\mathrm{Cor}} = 2.5 \times 10^{-9}\ \mathrm{m/s^2}$, for which we assume an initial cloud velocity error of $s_{v_{\mathrm{cloud,initial}}} = 10^{-7}\mathrm{m/s}$, a thermal cloud velocity of $v_{\mathrm{atom,therm}} = 10^{-6}\ \mathrm{m/s}$ and transverse cloud velocity of $v_{\mathrm{cloud}} = 2.5\mathrm{cm/s}$. For the sequential mode of operation, the cloud velocity is solely associated with thermal motion, and the Coriolis effects are limited to $s_{\mathrm{Cor}} = 2.0 \times 10^{-13}\mathrm{m/s^2}$, which includes the effect of the necessary lower sampling rate compared to the concurrent operational mode. Either case requires full attitude compensation with tilting mirrors, which reduce the satellite's attitude rates down to $\delta\omega = 5 \times 10^{-7}\mathrm{rad/s}$, assumed to be one order of magnitude worse than what the high-performance Astrix 200 laser gyroscope can measure (Section 2.1.2). Although such a CAI instrument is well suited as an accelerometer for ll-SST, it is unable to measure temporal gravity changes as a gradiometer since the sensitivity would be limited to $s_{V_{\mathrm{Cor}}} = 0.8$ mE for a distance between cloud-pairs (i.e., the gradiometer arm length) of $L_{GG} = 0.5$m.

We considered the errors related to the rotation of the measurements in the body (for gradiometry) or local (for ll-SST) frames to the Earth corotating frame (Section 4), which are of importance given the high accuracy of the measurements and the potential large acceleration or gravity gradient signal at LEO altitudes. These errors tie the signal amplitude of non-gravitational accelerations (for the case of accelerometers) or gravity gradient amplitudes (for gradiometry) with the attitude accuracy. They become important when the attitude measurements have limited accuracy, as is the case with classical attitude sensors. At the same time, the signal can be measured with increased accuracy, for example, with predicted quantum sensors. An example is the case of ll-SST equipped with a CAI accelerometer, Section 5.2.2, where the need for a 3D DFC system is required in the case the attitude derived with the DWS is retrieved with accuracy comparable to what GRACE-FO is capable of; if this can be done with an accuracy predicted for LISA, then no DFC is necessary.

In Section 5.2.2, we consider a CAI accelerometer with a noise level of $s_{\mathrm{CAI,ng}} = 3.2 \times 10^{-14}\mathrm{m/s^2}$ in the concurrent mode and $s_{\mathrm{CAI,ng}} = 1.1 \times 10^{-13}\mathrm{m/s^2}$ in the sequential mode, with the CAI parameters in Table 2, where notably the degree of entanglement is $\alpha = 0.25$, the number of atoms is $N = 10^7$, and the momentum space separation $\beta = 2$ is associated with a second kick. This example scenario intends to illustrate a possible path for the development of CAI instruments with increasing accuracy and their capabilities for gradiometry. In case of the instrument operating in sequential mode in an ll-SST formation, its sensitivity is not significantly limited by the effects of the Coriolis accelerations, since they have an amplitude of $s_{\mathrm{Cor}} = 1.2 \times 10^{-13}\ \mathrm{m/s^2}$. As a result, the LTI predicted for 2040 is not hampered in any way down to the sub-orbital frequencies. In contrast, in the case of ll-SST with an electrostatic accelerometer (Section 5.2.1), the MicroSTAR electrostatic accelerometer noise spectrum dominates over the LTI in all frequencies below 30 mHz. Of note is that in either case, the effect of errors in reconstructing the centrifugal accelerations acting on the frame connecting the two satellites (Section 3.1.1.1) is dominant over all instrument errors. We forward the hypothesis that the extrapolation of the spectra of this error into high frequencies may be inaccurate, because there is no way to measure it. We also postulate that advanced GNSS data processing strategies and exploiting multiple observation techniques (Ge et al., 2022; Li et al., 2023), may reduce the errors in the radial velocity, which is their main cause.

In Section 5.3, we apply the CAI parameters derived for the case of ll-SST to quantum gradiometry, using the CAI operating in sequential mode. We demonstrate that, unlike the case for ll-SST, this instrument is barely sensitive enough to resolve temporal changes in Earth's gravity. The maximum SH degree it is sensitive to is 17, and the Coriolis accelerations make it impossible to measure spatial features with a length associated with SH degree 2 or longer. This example illustrates the much-reduced gravity gradient signal amplitude, compared to gravitational accelerations, to which ll-SST is sensitive.

We recognise that there are important technological challenges associated with the solutions considered in this study. We have generally neglected such details because we restrict our analysis to the conceptual level. The obvious consequence is the high cost for the necessary technical and engineering developments, most notably those associated with i) the highly accurate rotation compensation provided by the tilting mirrors and ii) the cooling of the atom cloud during the BEC preparation that is required to reduce the thermal velocity of the atoms as assumed in this study. We also note that we did not model the loss of interferometric contrast associated with the scenarios without rotation compensation, since that parameter is specific to the design of the instrument.

Nevertheless, we have demonstrated that the effects of inaccurately measured attitude in the Coriolis accelerations are of paramount importance to the success of the CAI satellite gravimetry, even in the quiet environment of space. We emphasise that this is a problem that cannot be solved with novel attitude control strategies (Zinage et al., 2024) or eventual advanced ground processing techniques. There is a limitation in observing the attitude of the spacecraft with unprecedented accuracy, which is needed to ensure that the high quality of the data collected by the CAI instrument is fully exploited. Until this is done, we propose that it is significantly less technically challenging to implement gravimetric CAI technology in the context of the ll-SST measurement concepts. We also demonstrate that it is much easier to take advantage of the CAI accelerometer operating in sequential mode to circumvent the limitations of attitude data quality. We further contextualise that this conclusion is done a result of the limits of this study; therefore, any significant improvement in the attitude data quality, especially if gathered by the CAI instrument itself, would effectively unleash the full potential of CAI gradiometry, which, it should be emphasised, is not affected by the errors in reconstructing the centrifugal accelerations of the ll-SST baseline.

# 7. Acknowledgements

The work presented in this paper was conducted in the framework of the European Space Agency (ESA) study entitled *Quantum Space Gravimetry for monitoring Earth's Mass Transport Processes* (QSG4EMT). The QSG4EMT study has the objective to analyse QSG mission architectures that can optimally recover the time variable gravity field of the Earth, which relates to mass transport processes of interest for operational applications, as well as for Earth Sciences, and propose candidate QSG mission concepts with mission requirements that advance the science and satisfy the needs of the user community to a greater extent than the state-of-the-art and planned missions.
We thank Vitali Müller, from the Albert-Einstein-Institut in Hannover, for the insightful predictions of the LTI performance in the near future.

# 8. Appendix

## 8.1. Derivations

### 8.1.1. CAI accelerometer observation equation

Assume that the axis of the Raman laser is aligned with the i-axis:

$$\frac{\boldsymbol{k}_{\text{eff}}}{k_{\text{eff}}} = \boldsymbol{e}_i.$$

Starting from Equation ( 27 ):

$$\Phi = \boldsymbol{k}_{\text{eff}} \cdot (\boldsymbol{a}_{\text{cloud}} - \boldsymbol{a}_{\text{mirror}})T^2$$

The non-gravitation accelerations $\boldsymbol{a}_{\text{ng}} = \boldsymbol{a}_{\text{mirror}}$ project along $\boldsymbol{e}_i$ are given by:

$$\Phi = \boldsymbol{k}_{\text{eff}} \cdot \left(\boldsymbol{a}_{\text{cloud}} - \boldsymbol{a}_{\text{ng}}\right)T^2$$

$$\frac{\Phi}{k_{\text{eff}}} = \frac{\boldsymbol{k}_{\text{eff}}}{k_{\text{eff}}} \cdot \left(\boldsymbol{a}_{\text{cloud}} - \boldsymbol{a}_{\text{ng}}\right)T^2$$

$$\frac{\Phi}{k_{\text{eff}}T^2} = \boldsymbol{e}_i \cdot \left(\boldsymbol{a}_{\text{cloud}} - \boldsymbol{a}_{\text{ng}}\right) = \boldsymbol{e}_i \cdot \boldsymbol{a}_{\text{cloud}} - \boldsymbol{e}_i \cdot \boldsymbol{a}_{\text{ng}}$$

$$\boldsymbol{e}_i \cdot \boldsymbol{a}_{\text{ng}} = -\frac{\Phi}{k_{\text{eff}} T^2} + \boldsymbol{e}_i \cdot \boldsymbol{a}_{\text{cloud}}$$

Replacing the acceleration of the atom cloud is given by Equation ( 29 ) in the equation above, which results in Equation ( 33 ):

$$\boldsymbol{e}_i \cdot \boldsymbol{a}_{\text{ng}} = -\frac{\Phi}{k_{\text{eff}} T^2} + \boldsymbol{e}_i \cdot \left(-\left(V - \boldsymbol{\Omega}^2 - \dot{\boldsymbol{\Omega}}\right)(\boldsymbol{r}_{\text{cloud}} - \boldsymbol{r}_{\text{CoM}}) + 2\boldsymbol{\omega} \times \boldsymbol{v}_{\text{cloud}}\right)$$

### 8.1.2. CAI gradiometer observation equation

From Equation ( 49 ):

$$\frac{\Phi_{i,l}}{k_{\text{eff}} T^2} = \boldsymbol{e}_i \cdot \left(-\boldsymbol{a}_{\text{ng,l}} - \left(V - \boldsymbol{\Omega}^2 - \dot{\boldsymbol{\Omega}}\right)\left(\boldsymbol{r}_{\text{cloud,l}} - \boldsymbol{r}_{\text{CoM}}\right) + 2\boldsymbol{\omega} \times \boldsymbol{v}_{\text{cloud,l}}\right),$$

the phase measurement of cloud l is:

$$\Phi_{i,l} = k_{\text{eff}} T^2 \boldsymbol{e}_i \cdot \left(-\boldsymbol{a}_{\text{ng,l}} - \left(V - \boldsymbol{\Omega}^2 - \dot{\boldsymbol{\Omega}}\right)\left(\boldsymbol{r}_{\text{cloud,l}} - \boldsymbol{r}_{\text{CoM}}\right) + 2\boldsymbol{\omega} \times \boldsymbol{v}_{\text{cloud,l}}\right),$$

and the differential measurement $\delta\Phi_i$ is:

$$\delta\Phi_i \equiv \Phi_{i,1} - \Phi_{i,2} =$$
$$k_{\text{eff}} T^2 \boldsymbol{e}_i \cdot \left(-\boldsymbol{a}_{\text{ng,1}} - \left(V - \boldsymbol{\Omega}^2 - \dot{\boldsymbol{\Omega}}\right)\left(\boldsymbol{r}_{\text{cloud,1}} - \boldsymbol{r}_{\text{CoM}}\right) + 2\boldsymbol{\omega} \times \boldsymbol{v}_{\text{cloud,1}}\right) -$$
$$k_{\text{eff}} T^2 \boldsymbol{e}_i \cdot \left(-\boldsymbol{a}_{\text{ng,2}} - \left(V - \boldsymbol{\Omega}^2 - \dot{\boldsymbol{\Omega}}\right)\left(\boldsymbol{r}_{\text{cloud,2}} - \boldsymbol{r}_{\text{CoM}}\right) + 2\boldsymbol{\omega} \times \boldsymbol{v}_{\text{cloud,2}}\right)$$

If $\boldsymbol{r}$ is measured from the CoM, $\boldsymbol{v}_{\text{cloud,1}}$ and $\boldsymbol{v}_{\text{cloud,2}}$ are both given by $\boldsymbol{v}_{\text{cloud,therm}}$(assumed to be a random variable):

$$\frac{\delta\Phi_i}{k_{\text{eff}} T^2} = \boldsymbol{e}_i \cdot \left(-\boldsymbol{a}_{\text{ng,1}} + \boldsymbol{a}_{\text{ng,2}} - \left(V - \boldsymbol{\Omega}^2 - \dot{\boldsymbol{\Omega}}\right)\delta\boldsymbol{r}_{\text{cloud}} + 2\boldsymbol{\omega} \times \boldsymbol{v}_{\text{cloud,therm}}\right)$$

Since $\boldsymbol{a}_{\text{ng,1}} = \boldsymbol{a}_{\text{ng,2}}$ and $\delta\boldsymbol{r}_{\text{cloud}} = L_{\text{GG}}\boldsymbol{e}_i$, we arrive at Equation ( 51 ):

$$\frac{\delta\Phi_i}{k_{eff} T^2} = \boldsymbol{e}_i \cdot \left(-\left(\boldsymbol{V} - \boldsymbol{\Omega}^2 - \dot{\boldsymbol{\Omega}}\right)L_{\text{GG}}\boldsymbol{e}_i + 2\boldsymbol{\omega} \times \boldsymbol{v}_{\text{cloud,therm}}\right)$$

The scalar evaluation of this expression requires:

$$\boldsymbol{V}\boldsymbol{e}_i = \begin{bmatrix} V_{ii} \\ V_{ij} \\ V_{ik} \end{bmatrix} \text{ and } \boldsymbol{e}_i \cdot (\boldsymbol{V}\boldsymbol{e}_i) = V_{ii},$$

$$\boldsymbol{\Omega}^2\boldsymbol{e}_i = \begin{bmatrix} -\omega_j^2 - \omega_k^2 \\ \omega_i\omega_j \\ \omega_i\omega_k \end{bmatrix} \text{ and } \boldsymbol{e}_i \cdot (\boldsymbol{\Omega}^2\boldsymbol{e}_i) = -\omega_j^2 - \omega_k^2,$$

$$\dot{\boldsymbol{\Omega}}\boldsymbol{e}_i = \begin{bmatrix} 0 \\ \dot{\omega}_k \\ \dot{\omega}_j \end{bmatrix} \text{ and } \boldsymbol{e}_i \cdot \left(\dot{\boldsymbol{\Omega}}\boldsymbol{e}_i\right) = 0,$$

$$\boldsymbol{\omega} \times \boldsymbol{v} = \begin{bmatrix} \omega_j v_k - \omega_k v_j \\ \omega_k v_i - \omega_i v_k \\ \omega_i v_j - \omega_j v_i \end{bmatrix} \text{ and } \boldsymbol{e}_i \cdot (\boldsymbol{\omega} \times \boldsymbol{v}) = \omega_j v_k - \omega_k v_j,$$

resulting in:

$$\frac{\delta\Phi_i}{k_{eff} T^2} = -\left(V_{ii} + \omega_j^2 + \omega_k^2\right)L_{\text{GG}} + 2\omega_j v_{\text{cloud,therm},k} - 2\omega_k v_{\text{cloud,therm},j}$$

$$\frac{\delta\Phi_i}{k_{eff} T^2}\frac{1}{L_{\text{GG}}} = -V_{ii} - \omega_j^2 - \omega_k^2 + \frac{2}{L_{\text{GG}}}\omega_j v_{\text{cloud,therm},k} - \frac{2}{L_{\text{GG}}}\omega_k v_{\text{cloud,therm},j}$$

Rearranging, results in Equation ( 52 ):

$$V_{ii} = -\frac{1}{L_{\text{GG}}}\frac{\delta\Phi_i}{k_{eff} T^2} - \omega_j^2 - \omega_k^2 + \frac{2}{L_{\text{GG}}}\left(\omega_j v_{\text{cloud,therm},k} - \omega_k v_{\text{cloud,therm},j}\right)$$

For a CAI gradiometer oriented along the other axes:

$$V_{jj} = -\frac{1}{L_{\mathrm{GG}}}\frac{\delta\Phi_j}{k_{eff}T^2} - \omega_k^2 - \omega_i^2 + \frac{2}{L_{\mathrm{GG}}}\left(\omega_k v_{\mathrm{cloud,therm},i} - \omega_i v_{\mathrm{cloud,therm},k}\right),$$
$$V_{kk} = -\frac{1}{L_{\mathrm{GG}}}\frac{\delta\Phi_k}{k_{eff}T^2} - \omega_i^2 - \omega_j^2 + \frac{2}{L_{\mathrm{GG}}}\left(\omega_i v_{\mathrm{cloud,therm},j} - \omega_j v_{\mathrm{cloud,therm},i}\right),$$

derived considering the following relations:

$$\boldsymbol{e}_j \cdot \left(\boldsymbol{V}\boldsymbol{e}_j\right) = V_{jj} \text{ and } \boldsymbol{e}_k \cdot \left(\boldsymbol{V}\boldsymbol{e}_k\right) = V_{kk}$$
$$\boldsymbol{e}_j \cdot \left(\boldsymbol{\Omega}^2\boldsymbol{e}_j\right) = -\omega_i^2 - \omega_k^2 \text{ and } \boldsymbol{e}_k \cdot \left(\boldsymbol{\Omega}^2\boldsymbol{e}_k\right) = -\omega_i^2 - \omega_j^2$$
$$\boldsymbol{e}_j \cdot \left(\dot{\boldsymbol{\Omega}}\boldsymbol{e}_j\right) = \boldsymbol{e}_k \cdot \left(\dot{\boldsymbol{\Omega}}\boldsymbol{e}_k\right) = 0$$
$$\boldsymbol{e}_j \cdot (\boldsymbol{\omega} \times \boldsymbol{v}) = \omega_k v_i - \omega_i v_k \text{ and } \boldsymbol{e}_k \cdot (\boldsymbol{\omega} \times \boldsymbol{v}) = \omega_i v_j - \omega_j v_i$$

## 8.2. Angular Velocity Signal Magnitude

To assess the signal size of the angular velocity, we make several assumptions about the orbit and attitude control. First, we assume an orbit at an altitude of 500 km, which gives an orbital period of $T_{orb} = 95\ min$. Next, we assume that the satellite is nadir pointing, which results in a mean pitch rate of:

$$\mathrm{mean}\left(\omega_y\right) = 2\pi/T_{orb} = 1.1\ \mathrm{mrad/s}. \tag{79}$$

Further, we assume that the satellite is pointing in the direction of atmospheric flow to minimise the effects of drag. The direction of the flow relative to the satellite is composed of the inertial velocity of 7.6 km/s and the corotation of the atmosphere of 500 m/s at the equator in an eastward direction. The worst case is a polar orbit, in which the inertial velocity is perpendicular to the velocity of the corotating atmosphere. The maximum yaw angle at the equation is $\arcsin(0.5\ km/7.6\ km) = 3.77°$. Since the velocity of the corotating atmosphere is zero at the pole, the yaw angle is also zero. Thus, the yaw angle changes from 3.77° to 0° during a quarter of an orbit, i.e., by $3.77°/(T_{orb}/4) = 4.6 \times 10^{-5} rad/s$. To calculate the RMS of the angular velocity, we assume that the yaw angle varies like a sine function with an orbital period, i.e.:

$$\psi = 3.77° \sin\left(2\pi t/T_{orb}\right). \tag{80}$$

The angular velocity is the time derivative of that function:

$$\omega_z = \frac{\partial\psi}{\partial t} = 3.77°\ 2\pi/T_{orb} \cos\left(2\pi t/T_{orb}\right). \tag{81}$$

The integral of the squared function is:

$$\begin{aligned}
\int_0^{T_{orb}} \omega_z\, dt &= \left(3.77°\ 2\pi/T_{orb}\right)^2 \int_0^{T_{orb}} \cos^2\left(2\pi t/T_{orb}\right) dt \\
&= \left(3.77°\ 2\pi/T_{orb}\right)^2 \frac{1}{2}\int_0^{T_{orb}} 1 - \sin\left(4\pi t/T_{orb}\right) dt \\
&= \left(3.77°\ 2\pi/T_{orb}\right)^2 \frac{1}{2}\left[t - 4\pi/T_{orb}\cos(4\pi t/T_{orb})\right]_0^{T_{orb}} \\
&= \left(3.77°\ 2\pi/T_{orb}\right)^2 T_{orb}/2.
\end{aligned} \tag{82}$$

The RMS is then the square root of the integral divided by $T_{orb}$:

$$\mathrm{RMS}(\omega_z) = \sqrt{\frac{1}{T_{orb}}\int_0^{T_{orb}} \omega_z\, dt} = 3.77° \frac{2\pi}{T_{orb}\sqrt{2}} = 5.1 \times 10^{-5}\ rad/s. \tag{83}$$

When magnetic torquers are the only means for attitude control, there is typically no control of the roll at the equator because the magnetic field lines are parallel to the roll axis. With that in mind, we assume that the RMS of the roll rate is:

$$\mathrm{RMS}(\omega_x) = 0.1\ \mathrm{mrad/s}.$$

## 8.3. Non-gravitational acceleration signal magnitude

Some of the error propagations require assumptions about the non-gravitational acceleration signal size. Considering that aerodynamic accelerations are large at low altitudes, we use GOCE data as a worst-case scenario. Typical acceleration signal sizes are reported in Table 3.

Table 3: Non-gravitational signal size in nm/s$^2$, mean for long-track and standard deviation for cross-track and radial directions, for the case of the GOCE mission, considering 1D and (hypothetical) 3D drag control (Visser and van den IJssel, 2016).

| [nm/s$^2$] | GOCE 230km | 1D Drag Control | 3D Drag Control |
|---|---|---|---|
| Along-track (mean) | 10000 | 10 | 10 |
| Cross-track (STD) | 289 | 289 | 10 |
| Radial (STD) | 22 | 22 | 10 |

The DFC of GOCE was estimated to reduce non-gravitation accelerations down to 1 $\mathrm{nm/s^2}$ (Christophe et al., 2018). As such, the assumptions on 1D and 3D DFC shown in Table 3 are conservative.

## 8.4. Gravity gradient signal magnitude

Assuming a LEO orbit, the signal amplitude of the (symmetric) gravity gradient tensor in the LHRF $\boldsymbol{V}^{(\mathrm{LHRF})}$ at 450km altitude is (e.g. Rosen 2021):

$$\mathrm{RMS}\left(\boldsymbol{V}^{(\mathrm{LHRF})}\right) \approx \begin{bmatrix} 1200 & 0 & 0 \\ & 1200 & 0 \\ & & -2400 \end{bmatrix} + \begin{bmatrix} 3.01 & 0.03 & 0.09 \\ & 4.22 & 4.38 \\ & & -7.23 \end{bmatrix} [E].$$

The first term is associated with the signal caused by the central term of the gravity field, and the static gravity field of the Earth causes the second term. We ignored the term caused by the temporal variations of the gravity field.

## 8.5. Reference frames

### 8.5.1. Earth-centred, Earth-fixed reference frame

The ECEF reference frame, also known as Terrestrial Reference Frame (TRF), Earth Centred Fixed reference frame (ECF), Conventional Terrestrial Frame (CTF) and Earth-Fixed Reference Frame (EFRF), is centred at the COM of the Earth. The x-axis points towards longitude 0 degrees and is parallel to the equatorial plane. The y-axis, also parallel to the equatorial plane, points towards 90 degrees longitude. The z-axis points towards the North pole. The TRF is suitable for representing geophysical processes in the Earth system.

### 8.5.2. Earth-centred, inertial reference frame

The ECI reference frame, also know as Celestial Reference Frame (CRF), Conventional Inertial Frame (CIF) and Inertial Reference Frame (IRF), is the quasi-inertial reference frame defined with the x-axis pointing towards the direction of the mean equinox at J2000.0 epoch (noon on 1st of January, 2000, Terrestrial Time), the z-axis points to the North pole at that epoch (formally, orthogonal to the plane defined by the mean equator at J2000.0) and the y-axis closes the right-handed set of orthogonal axes. The origin of the CRF is located at the COM of the Earth. The quasi-inertial property of this reference frame makes it suitable to represent the motion of a spacecraft.

### 8.5.3. Satellite Reference Frame

The SRF is used to define the geometry of the spacecraft. The axes are connected to the body of the spacecraft in a predefined (but otherwise arbitrary) way. The attitude determination system provides measurements that connect the inertial to this reference frame. The pressure forces due to Solar and Earth radiation, as well as those of aerodynamic nature, are most suitably defined in this reference frame since the orientation of the satellite exterior panels is constant. This reference frame also serves as a basis to define the reference frame of the various sensors and actuators on board the satellites, such as the accelerometer(s), the star-tracker(s), magneto-torquers, propulsion, etc.

## 8.6. Error in reconstructing the centrifugal accelerations in ll-SST

In this Section, we present our assumptions for the parameters in Eq. ( 26 ). One option to derive representative amplitude and spectra of the errors in the radial velocity $v_z$ is compute the difference between 1) the inter-satellite velocities computed from orbit integration and 2) the GRACE's KBR data. This was done for January, February, July, August and September of 2006, with a total of over 6 billion data points (Teixeira da Encarnação, 2015). This analysis removed data that had low SNR, surrounding gaps resulting from phase breaks, discontinuities in the integrated orbit arcs (every 6 hours) and outliers, which were removed iteratively when larger than 4-$\sigma$, until changes in the STD of the remaining data did not change significantly.

The data considered in the analysis revealed an STD of 2.19 $\mu m/s$, which is close to the 4 $\mu m/s$ that Ditmar et al. (2012) considered in their study, when a less strict data pre-processing scheme was considered.

Importantly, we assume in this study that the spectra of these errors are a simplified representation of the results in Figure 6.3 of Teixeira da Encarnação (2015), as shown in Figure 21 and reported numerically in Table 4.

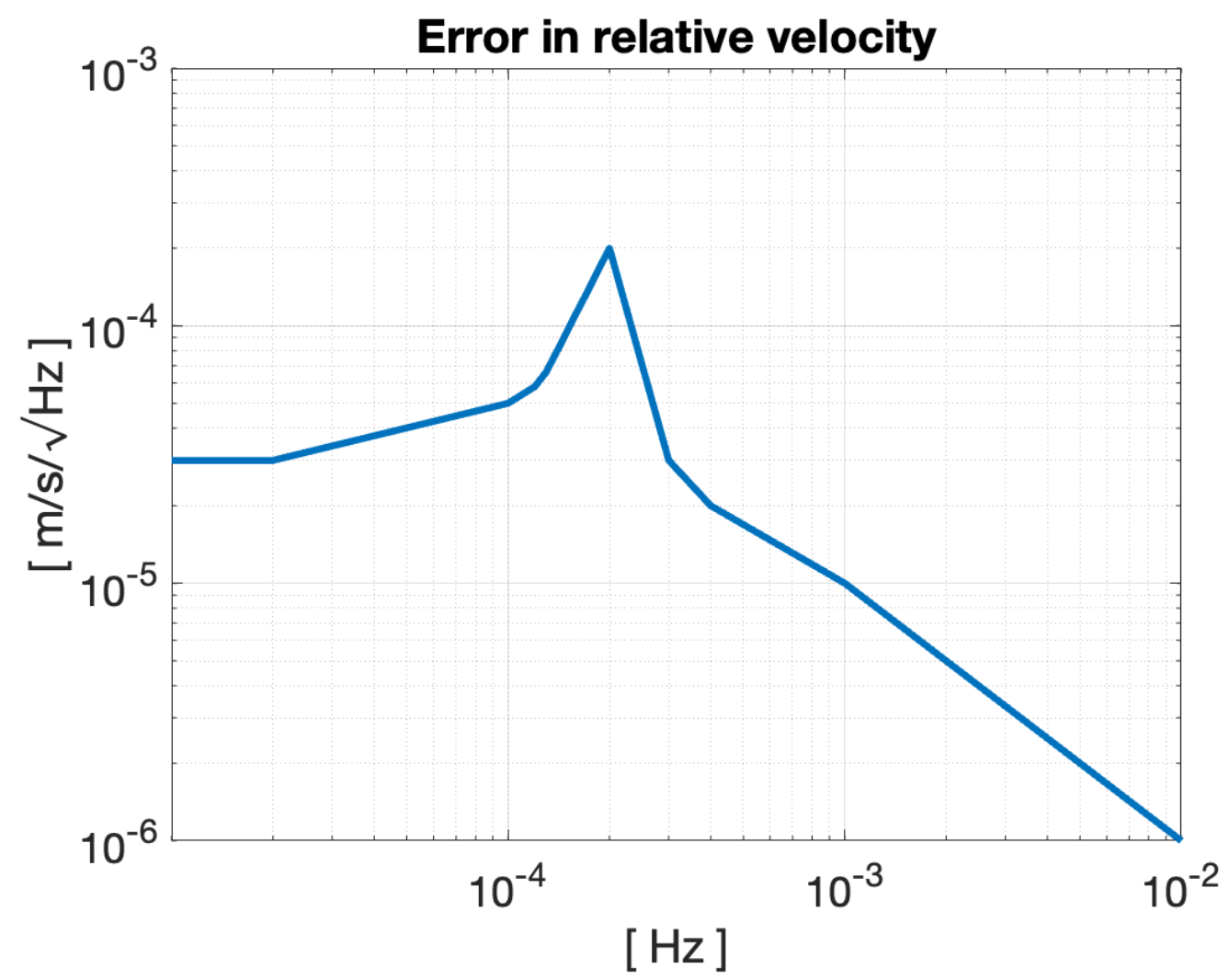

Figure 21: Errors in the relative velocity, estimated from the difference between KBR data and the distance computed from integrated orbits.

Table 4: Numerical values of the assumed amplitude spectra of the radial velocity errors

| Frequency [mHz] | 0.01 | 0.02 | 0.1 | 0.125 | 0.2 | 0.3 | 0.4 | 1 | 10 |
|---|---|---|---|---|---|---|---|---|---|
| Amplitude [m/s/√Hz] | $3 \times 10^{-5}$ | $3 \times 10^{-5}$ | $5 \times 10^{-5}$ | $6 \times 10^{-5}$ | $2 \times 10^{-4}$ | $3 \times 10^{-5}$ | $2 \times 10^{-5}$ | $1 \times 10^{-5}$ | $1 \times 10^{-5}$ |

We further consider that the velocity errors along the LoS direction are representative in magnitude and spectra of the radial velocity errors, which is relevant to computing the error in reconstructing the centrifugal accelerations. It is important to have a good estimate of the spectra of radial velocity errors; for example, if white noise is considered, their propagation to the range-acceleration space would incorrectly magnify the high frequencies.

## 8.7. Methodology for Selection of CAI Parameters

Our objective is to identify instrument configurations that deliver acceleration sensitivity suitable for low-low SST missions while remaining technically plausible for a flight implementation. To do this, we coupled a physics-based performance model to a set of empirically-derived cost proxies that penalise design choices which, in the authors' experience, drive mass, volume, complexity and therefore programme risk. The section below summarises the two building blocks of the trade-space exploration and the way they are combined.

### 8.7.1. Performance metric

Cold-atom interferometer (CAI) noise is dominated by quantum-projection (shot) noise in the measurement of the transition probability. Assuming technical noise is kept sub-dominant, we can use Equation ( 20 ) to calculate the performance in terms of acceleration sensitivity. We use this equation also to define the levers available to the instrument designer: longer interferometer time, larger momentum transfer, more atoms and stronger entanglement all lower the noise floor, but each lever has engineering consequences that quickly become mission-limiting in a spaceborne context

### 8.7.2. Engineering cost proxies

Four parameters were selected as the dominant cost drivers:

Table 5: Cost Drivers and Proxy Variables

| **Symbol** | **Physical lever** | **Rationale for penalising** | **Proxy variable $x$** | **Shaping Factor $q$** | **Weight $w$** |
|---|---|---|---|---|---|
| $T$ | Pulse-separation time | System isolation, mechanical and optical stability | $T(s)$ | 0.01 | 0.1 |
| $N$ | Atom number | Laser power & linewidth, cooling/launch duty-cycle, vacuum quality, trap quality | $\log_{10} N$ | 1 | 1 |
| $\alpha$ | Entanglement degree | control systems, sensitivity to decoherence, high-finesse optics | $\alpha$ | 2 | 5 |
| $H$ | Chamber height | Structural mass, magnetic shielding, field homogeneity, alignment precision, satellite real estate | $H = vT$ with $v = \frac{\beta}{2}\frac{\hbar k}{m}$ | 10 | 10 |

The chamber height $H$ is evaluated from the cloud's ballistic half-excursion $vT$; for the $Rb^{87}$ Raman, this gives $v \approx 3\beta$ mm/s.
Each proxy is mapped to a dimensionless cost coefficient:

$$C = \frac{e^{qx} - 1}{e^{qx_{\max}} - 1}$$

which monotonically scales from 0 at the reference value ($x = 0$) to 1 at an upper admissible bound $x_{\max}$. The shaping factor $q$ allows us to encode how abruptly cost is believed to rise as the limit is approached; steep curves (large $q$) mimic "cliff edges" in manufacturability, while shallow curves reflect gradual cost growth.
Finally, subjective weights $w_i$ convert the individual coefficients into a composite score

$$C_{\text{total}} = \sum_i w_i \, C_i$$

The large weight on vacuum-chamber height echoes launch-vehicle and thus satellite volume constraints and centre-of-mass offsets that quickly dominate spacecraft design. Entanglement complexity is also strongly penalised because only modest levels of spin-squeezing have yet been demonstrated with a moderate number of atoms (Greve et al., 2022).

### 8.7.3. Trade-space exploration

For a grid of $(\beta, T, N, \alpha)$, we compute the sensitivity $s_{\text{CAI,ng}}^{(s2s)}$ and total cost above. Parameter combinations that satisfy a notional noise requirement are plotted in cost–performance space and ranked by increasing cost, providing a means to identify configurations that offer relatively favourable trade-offs. The resulting distribution of design points is illustrated in Figure 8, with sensitivity on the x-axis, cost on the y-axis, and each point representing a distinct parameter set. The purpose of this approach is not to prescribe an optimal instrument design, but to establish a transparent and adaptable framework for first-order trade-space exploration. The performance model and cost functions are constructed to be analytically simple and empirically tunable, allowing for updates as new experimental results and system-level constraints emerge. In this way, the methodology facilitates early-stage assessment of design trends and sensitivities, supporting the iterative development of CAI concepts for spaceborne gravity missions.

## Abbreviations and Acronyms

| | |
|---|---|
| ASD | Amplitude Spectral Density |
| BEC | Bose-Einstein Condensate |
| CAI | Cold Atom Interferometer/interferometry |
| CoM | Centre of Mass |
| DKC | Delta Kick-Collimation |
| DFC | Drag-Free Control |
| DWS | Differential wavefront sensor |
| ECEF | Earth-centred, Earth-fixed (reference frame) |
| ECI | Earth-centred, inertial (reference frame) |
| ESA | European Space Agency |
| GG | Gravity Gradiometry/gradiometer |
| GOCE | Gravity field and steady-state Ocean Circulation Explorer |
| GRACE | Gravity Recovery and Climate Experiment |
| GRACE-FO | GRACE Follow-On |
| IMU | Inertial Measurement Unit |
| ICE | Interférométrie atomique à sources Cohérentes pour l'Espace |
| ISR | Inter-satellite range/ranging |
| KBR | K-Band Ranging |
| LEO | Low-Earth Orbit |
| LOS | Line-Of-Sight |
| LRI | Laser ranging interferometer |
| ll-SST | low-low Satellite-to-Satellite Tracking |
| MAIUS | Matter-Wave Interferometry in Weightlessness |
| μASC | Micro Advanced Stellar Compass |
| MOT | Magneto-Optical Trap |
| NGGM | Next Generation Gravity Mission |
| QSG | Quantum Space Gravimetry |
| QSG4EMT | QSG for monitoring Earth Mass Transport processes |
| PSD | Power Spectral Density |
| RMS | Root Mean Square |
| SH | Spherical Harmonic |
| SRF | Satellite Reference Frame |
| TMA | Triple Mirror Assembly |